\documentclass{aa}

\usepackage{graphicx}
\usepackage{supertabular}
\usepackage{epsfig}
\usepackage{natbib}
\usepackage[varg]{txfonts}
\usepackage{hyperref}
\usepackage[normalem]{ulem}

\newcommand{\logg}{\ensuremath{\log g}}

\newcommand{\mlp}{\ensuremath{\alpha_{\mathrm{MLT}}}}

\newcommand{\Teff}{\ensuremath{T_{\mathrm{eff}}}}

\newcommand{\beq}{\begin{equation}}
\newcommand{\eeq}{\end{equation}}

\newcommand{\xtmean}[1]{\ensuremath{\left\langle #1\right\rangle}}

\newcommand{\OoFe}{\ensuremath{\left[\mathrm{O}/\mathrm{Fe}\right]}}
\newcommand{\FeoH}{\ensuremath{\left[\mathrm{Fe}/\mathrm{H}\right]}}
\newcommand{\aoFe}{\ensuremath{\left[\mathrm{\alpha}/\mathrm{Fe}\right]}}

\newcommand{\logAO}{\ensuremath{A{\rm (O)}}}

\newcommand{\logAOOD}{\ensuremath{A{\rm (O)_{1D~LTE}}}}
\newcommand{\logAOTD}{\ensuremath{A{\rm (O)_{3D~LTE}}}}

\newcommand{\chisq}{\ensuremath{{\rm \chi^{2}}}}
\newcommand{\vmic}{\ensuremath{{\xi_{\mathrm{mic}}}}}

\newcommand{\vbrd}{\ensuremath{{v_{\mathrm{brd}}}}}
\newcommand{\Fobs}{\ensuremath{{F_{\mathrm{obs}}}}}
\newcommand{\Fsyn}{\ensuremath{{F_{\mathrm{syn}}}}}
\newcommand{\Wi}{\ensuremath{W_{\mathrm{i}}}}
\newcommand{\COBOLD}{{\tt CO$^5$BOLD}}
\newcommand{\LHD}{{\tt LHD}}

\newcommand{\ATLAS}{{\tt ATLAS9}}

\newcommand{\SYNTHE}{{\tt SYNTHE}}
\newcommand{\MARCS}{{\tt MARCS}}

\newcommand{\LINFOR}{{\tt Linfor3D}}

\begin{document}

\title{Three-dimensional hydrodynamical \COBOLD\ model atmospheres of red giant stars\\}
\subtitle{V. Oxygen abundance in the metal-poor giant HD~122563 from OH UV lines}

\author{
	    D.\,Prakapavi\v{c}ius\inst{1}
   \and A.\,Ku\v{c}inskas\inst{1,2}
   \and V.\,Dobrovolskas\inst{1}
   \and J.\,Klevas\inst{1}
   \and M.\,Steffen\inst{3,4}
   \and P.\,Bonifacio\inst{4}
   \and H.-G.\,Ludwig\inst{4,5}
   \and M.\,Spite\inst{4}
   }

\institute{
       Institute of Theoretical Physics and Astronomy, Vilnius University, Saul\.{e}tekio al. 5, Vilnius LT-10221, Lithuania \\
       \email{dainius.prakapavicius@tfai.vu.lt}
	   \and 
 	   Astronomical Observatory, Vilnius University, M. K. \v{C}iurlionio 29, Vilnius LT-03100, Lithuania
	   \and 
	   Leibniz-Institut f\"ur Astrophysik Potsdam, An der Sternwarte 16, D-14482 Potsdam, Germany
	   \and 
	   GEPI, Observatoire de Paris, CNRS, Universit\'{e} Paris Diderot, Place Jules Janssen, 92190 Meudon, France
	   \and
	   Landessternwarte -- Zentrum f\"ur Astronomie der Universit\"at Heidelberg, K\"onigstuhl 12, D-69117 Heidelberg, Germany
   }

\date{Received XXX; accepted XXX}

\abstract
 {Although oxygen is an important tracer of the early Galactic evolution, its abundance trends with metallicity are still relatively poorly known at $\FeoH \lesssim -2.5$. This is in part due to a lack of reliable oxygen abundance indicators in the metal-poor stars, in part due to shortcomings in 1D~LTE abundance analyses where different abundance indicators, such as OH lines located in the UV and IR or the forbidden [O~I] line at 630\,nm, frequently provide inconsistent results.}
   {In this study we determined the oxygen abundance in the metal-poor halo giant HD~122563 using a 3D hydrodynamical \COBOLD\ model atmosphere. Our main goal was to understand whether a 3D~LTE analysis may help to improve the reliability of oxygen abundances determined from OH UV lines in comparison to those obtained using standard 1D~LTE methodology.}
   {The oxygen abundance in HD~122563 was determined using 71 OH UV lines located in the wavelength range between $308-330$\,nm. The analysis was done using a high-resolution VLT~UVES spectrum with a 1D~LTE spectral line synthesis performed using the \SYNTHE\ package and classical \ATLAS\ model atmosphere. Subsequently, a 3D hydrodynamical \COBOLD, and 1D hydrostatic \LHD\ model atmospheres were used in order to compute 3D--1D abundance corrections. For this, the microturbulence velocity used with the 1D \LHD\ model atmosphere was derived from the hydrodynamical \COBOLD\ model atmosphere of HD~122563. The obtained abundance corrections were then applied to determine 3D~LTE oxygen abundances from each individual OH UV line.}
   {As in previous studies, we found trends of the 1D~LTE oxygen abundances determined from OH UV lines with line parameters, such as the line excitation potential, $\chi$, and the line equivalent width, $W$. These trends become significantly less pronounced in 3D~LTE. Using OH UV lines we determined a 3D~LTE oxygen abundance in HD~122563 of $A{\rm (O)_{\rm 3D~LTE}}=6.23\pm0.13$ ($\OoFe=0.07\pm0.13$). This is in fair agreement with the oxygen abundance obtained from OH IR lines, $A{\rm (O)_{\rm 3D~LTE}}=6.39\pm0.11$ ($\OoFe=0.23\pm0.11$), but it is noticeably lower than that determined using the forbidden [\ion{O}{i}] line, $A{\rm (O)_{\rm 3D~LTE}}=6.53\pm0.15$ ($\OoFe=0.37\pm0.15$). While the exact cause for this discrepancy remains unclear, it is very likely that non-LTE effects may play a decisive role here. Oxygen-to-iron ratios determined in HD~122563 using OH UV/IR lines and the forbidden [\ion{O}{i}] line fall on the lower boundary of the \OoFe\ distribution observed in the Galactic field stars at this metallicity and suggest a very mild oxygen overabundance with respect to iron, $\OoFe \lesssim 0.4$.}
   {}


\authorrunning{Prakapavi\v{c}ius et al.}
\titlerunning{Oxygen abundance in giants from UV OH lines}

\maketitle

%-------------------------------------------------------------------------
\section{Introduction}

   Oxygen is an important tracer of the chemical evolution of stellar populations. During the early stages of the formation of the Galaxy, oxygen was synthesized in massive stars and released into the interstellar medium on timescales of $10^6-10^7$ years. Since the amount of oxygen produced in a given stellar population depends on the details of its star formation history (such as initial mass function and star formation rate), the knowledge of the oxygen abundance in the oldest (i.e., metal-poor) Galactic stars may allow us to put stricter constraints on the possible scenarios of early Galactic evolution.
      
   Measurements of the oxygen abundance in metal-poor stars are difficult, for several reasons. First, only a few spectral lines of atomic oxygen are available for oxygen diagnostics in metal-poor ($\FeoH<-2$) stars: the forbidden [\ion{O}{i}] 630\,nm line in giants, and the permitted \ion{O}{i} 777\,nm triplet in main sequence stars and subgiants. Therefore, different spectral lines are used to study different types of stars. Besides, all atomic oxygen lines become very weak at these low metallicities - and thus difficult to measure. Because of these reasons, OH vibrational-rotational and pure rotational lines in the infrared ($\sim1500-2100$\,nm) and electronic lines in the ultraviolet ($\sim310-330$\,nm) are often used to probe the oxygen abundance in metal-poor stars. However, oxygen abundances obtained from OH lines typically show large line-to-line scatter and/or trends with the line equivalent width and/or excitation potential \citep[see, e.g.,][]{MB02, BMS03,Aoki15}. Finally, oxygen abundances obtained using different indicators are frequently inconsistent. For example, abundances determined using OH lines may be up to $\sim 0.4$\,dex higher than those obtained using the [\ion{O}{i}] line \citep[see, e.g.,][and references therein]{DKB15}. All these factors lead to large systematic uncertainties in the derived oxygen abundances. As a consequence, our current knowledge about the oxygen trends at low metallicities is still incomplete: some authors claim that the \OoFe\ ratio increases towards the lowest metallicities and reaches $\OoFe \approx\ 1.0$ at $\FeoH\approx -3$ \citep[e.g.,][]{IRG01}, while others find that there is a plateau around $\OoFe \approx\ 0.4-0.7$ at $\FeoH\approx -2 \ldots -3$ \citep[e.g.,][]{CDS04, AAC15}.
      
   It is possible that inconsistencies in the oxygen abundances obtained using different spectral indicators are partly caused by insufficient physical realism of the classical abundance analysis techniques which rely on 1D hydrostatic model atmospheres. It is well known that oxygen lines are sensitive to 3D hydrodynamical effects, such as convection and/or shock waves \citep[e.g.,][]{CAT07,KSL13,DKS13}. However, these effects are not properly taken into account with the classical approach. Although the influence of 3D hydrodynamical effects on the spectral line strengths is different for different lines and depends on the structure of a given stellar atmosphere (which, in turn, is defined by the atmospheric parameters of a given star), 3D--1D differences in the oxygen abundances can reach or even exceed $\sim0.4$\,dex \citep[][]{CAT07,DKS13}. Such differences are similar in their size to the discrepancies in oxygen abundances obtained using different indicators described above and therefore may point to inadequacies in 1D model atmospheres as a possible culprit\footnote{Although non-local thermodynamic equilibrium (NLTE) effects may play an important role here too, their study is beyond the scope of the present work.}.
      
   In \citet[][]{DKB15} we therefore investigated whether 3D hydrodynamical model atmospheres may help to solve these problems. In that study we determined 3D~LTE oxygen abundances in four metal-poor red giants using infrared (IR) vibrational-rotational OH lines. We found that in the 3D~LTE analysis the difference between the average oxygen abundances obtained in the four metal-poor giants using OH IR lines and [\ion{O}{i}] 630\,nm line was $0.09$\,dex, down from $0.34$\,dex determined in 1D~LTE\footnote{It is known that the forbidden [\ion{O}{i}] line is insensitive to 3D hydrodynamical and NLTE effects, therefore oxygen abundances determined using this line offer a convenient reference point \citep[see, e.g.,][and references therein]{DKB15}.}. Unfortunately, it is still unclear whether similar improvements could be expected in case of OH lines that are located in the ultraviolet (UV) part of the spectrum. To our knowledge, the only studies where OH UV lines were investigated using 3D hydrodynamical model atmospheres were carried out by \citet{AG01}, who focused on dwarfs, \citet[][]{JGH10, BBL10} where the authors targeted a sample of metal poor subgiant stars; and by \citet{BCK15} who studied the extremely metal-poor halo subgiant SMSS J031300.36-670839.3 (though the atmospheric parameters of this star, $\Teff=5125$\,K and $\log g=2.3$, are in fact more similar to those of a giant) and \citet{CAT06}, who analysed a giant and a dwarf. Other studies of red giant stars are still lacking.
      
   In the present paper we therefore extend our previous work and focus on OH UV lines which we use to determine 1D~LTE and 3D~LTE oxygen abundances in the well-studied metal-poor giant HD~122563. The main goals of this work are: (a) to better understand the role of 3D hydrodynamical effects/convection in the formation of OH UV lines in the atmospheres of metal-poor red giant stars; (b) to find out whether the analysis of OH UV lines with 3D hydrodynamical model atmospheres may provide oxygen abundances that are more precise and reliable than those obtainable in 1D~LTE analysis; and (c) to determine whether the use of 3D hydrodynamical model atmospheres may help to reconcile abundances determined using OH UV, OH IR, and [\ion{O}{i}] lines. Therefore, ultimately, we aim to provide a prescription for future studies of oxygen abundances based on 3D~LTE analysis of OH UV lines that, besides of being stronger and more numerous than atomic oxygen lines, frequently can be the only available indicators of oxygen abundance in the metal-poor red giants.
      
   The paper is structured as follows: Our target object, i.e., the metal-poor red giant HD~122563, the spectroscopic material, the 1D hydrostatic and 3D hydrodynamical model atmospheres, and the 1D and 3D abundance analysis techniques are described in Sect.~\ref{sect:method}. The obtained results are presented and discussed in Sect.~\ref{sect:results}, while the main findings and conclusions are summarized in Sect~\ref{sect:conclusions}.
        
%-------------------------------------------------------------------------
\section{Methodology}\label{sect:method}

   In order to determine oxygen abundances in the metal-poor red giant HD~122563 from OH UV lines, we used classical 1D hydrostatic and 3D hydrodynamical model atmospheres together with 1D~LTE and 3D~LTE spectral synthesis techniques. In what follows below we briefly describe our motivation for selecting HD~122563 as the target for this pilot study, our choice of spectroscopic observational material, the model atmospheres and spectral synthesis techniques, and the methodology that we used to determine 1D~LTE and 3D~LTE oxygen abundances and their uncertainties in HD~122563.
   
\subsection{Target object: HD~122563}\label{sect:hd122563}

   In this work we focus on a well-studied metal-poor red giant HD~122563. This bright ($V = 6.2$) halo star has been thoroughly investigated in a number of studies and has reliably determined atmospheric parameters, as well as abundant high quality spectroscopic observations. All this makes HD~122563 a suitable target for the study of OH UV line formation in the atmospheres of metal-poor red giant stars. 
   
   After assessing available determinations of the atmospheric parameters of HD~122563, we have chosen to use $\Teff = 4600$\,K and $\log g = 1.60$, as determined using precise interferometric measurements by \citet{CTB12}. The metallicity, $\FeoH = -2.60$, was adopted from \citet{MGS11}. This result was obtained using 1D~NLTE abundance analysis techniques and the model atmosphere characterized by the atmospheric parameters identical to those used in our study (more recently \citealt{JHS14} has obtained a similar value of $\FeoH = -2.64$ which was determined in a careful 1D~NLTE analysis of \ion{Fe}{i} and \ion{Fe}{ii} lines in HD~122563).

\subsection{The observed spectrum of HD~122563}\label{sect:obs}

   We used a high-resolution (R = 60\,000) spectrum of HD~122563 that was obtained with the UVES spectrograph mounted on the ESO VLT UT2 telescope. A reduced spectrum was taken from the ESO UVES Paranal Observatory Project (UVES-POP) archive \citep[program ID 266.D-5655;][]{BJL03}. The UVES-POP spectrum of HD~122563 was obtained using both the blue and red arm of the spectrograph, and covered a total spectral range of $304-1040$\,nm. In our study we focused only on the UV part of the spectrum where OH UV lines used for the oxygen abundance analysis were located, i.e., the region between $308-330$\,nm. In the observed spectrum the continuum S/N ratio in this wavelength range varied from $\sim55@310$\,nm to $\sim195@330$\,nm (see Appendix~\ref{SN-ratio} for details related to the determination of S/N).

\begin{table}[]
	\centering
	\caption{Atmospheric parameters of the model atmospheres used in this work. In case of the 3D model atmosphere temporal RMS of the \Teff\ is also noted.}\label{tab:models}
	
	\begin{tabular}{|c|c|c|c|}
		\hline 
		Model          & \Teff         & \logg    & \FeoH    \\ 
		atmosphere     & K             & [cgs]    & dex       \\ 
		\hline 
		\ATLAS\ (1D)   & 4600          & 1.6      & $-2.60$ \\ 
		\LHD\ (1D)     & 4600          & 1.6      & $-2.50$ \\
		\COBOLD (3D)   & $4597 \pm 7 $ & 1.6      & $-2.50$ \\ 
		\hline 
	\end{tabular} 
\end{table}

\subsection{Model atmospheres}\label{sect:models}

   Three types of model atmospheres were used in this work: 
   \begin{list}{$\bullet$}{}
      \item a 1D hydrostatic \ATLAS\ model atmosphere;
      \item a 1D hydrostatic model atmosphere computed using the \LHD\ code \citep{CLS08};
      \item a 3D hydrodynamic model atmosphere computed with the \COBOLD\ package \citep{FSL12}.
   \end{list}

   \noindent Both 1D model atmospheres were identical to those used in \citet{DKB15} while our 3D model atmosphere had the same atmospheric parameters but was more extended in the vertical direction than the one used in \citet[][see below]{DKB15}. The atmospheric parameters of the model atmospheres are listed in Table~\ref{tab:models}.
   
   The three model atmospheres were computed having slightly different purposes in mind. The classical \ATLAS\ model atmosphere was used to determine oxygen abundances from the OH UV lines in the observed spectrum of HD~122563 (see Sect.~\ref{sect:1D_abund_method}). The model atmosphere was computed using the Linux port of the \ATLAS\ code \citep{SBC04,Sbor05}. Model computations were done with the overshooting switched off and the mixing length parameter set to $\mlp = 1.25$. The radiative transfer included continuum scattering and was solved by adopting NEWODF opacity distribution functions (ODFs) that were computed with metallicity $\FeoH = -2.5$ and microturbulence velocity of $\xi = 2$\,km\,s$^{-1}$ \citep{CK03}.
   
   The 1D hydrostatic \LHD\ and 3D hydrodynamical \COBOLD\ model atmospheres were used for computing 3D--1D LTE abundance corrections which were further used to determine 3D~LTE oxygen abundances (see Sect.~\ref{sect:3D_abund_method}). The abundance corrections were also used to assess the influence of convection on the formation of OH UV lines (Sect.~\ref{sect:results}). The \LHD\ and \COBOLD\ model atmospheres were computed using identical atmospheric parameters, chemical composition, equation of state, opacities, and radiative transfer scheme. Our goal therefore was to minimize the differences between the \LHD\ and \COBOLD\ models so that the role of convection in the OH UV line formation could be assessed by comparing the differences in the atmospheric structures and observable properties of the two model atmospheres (see Sect.~\ref{sect:results})\footnote{Such differential approach has been used in our earlier studies as well; see, e.g., \citet{DKS13}, \citet{KSL13}, \citet{DKB15}, for more details on this methodology.}.

   \begin{figure}[!t]
   	\centering
   	\includegraphics[width=9cm]{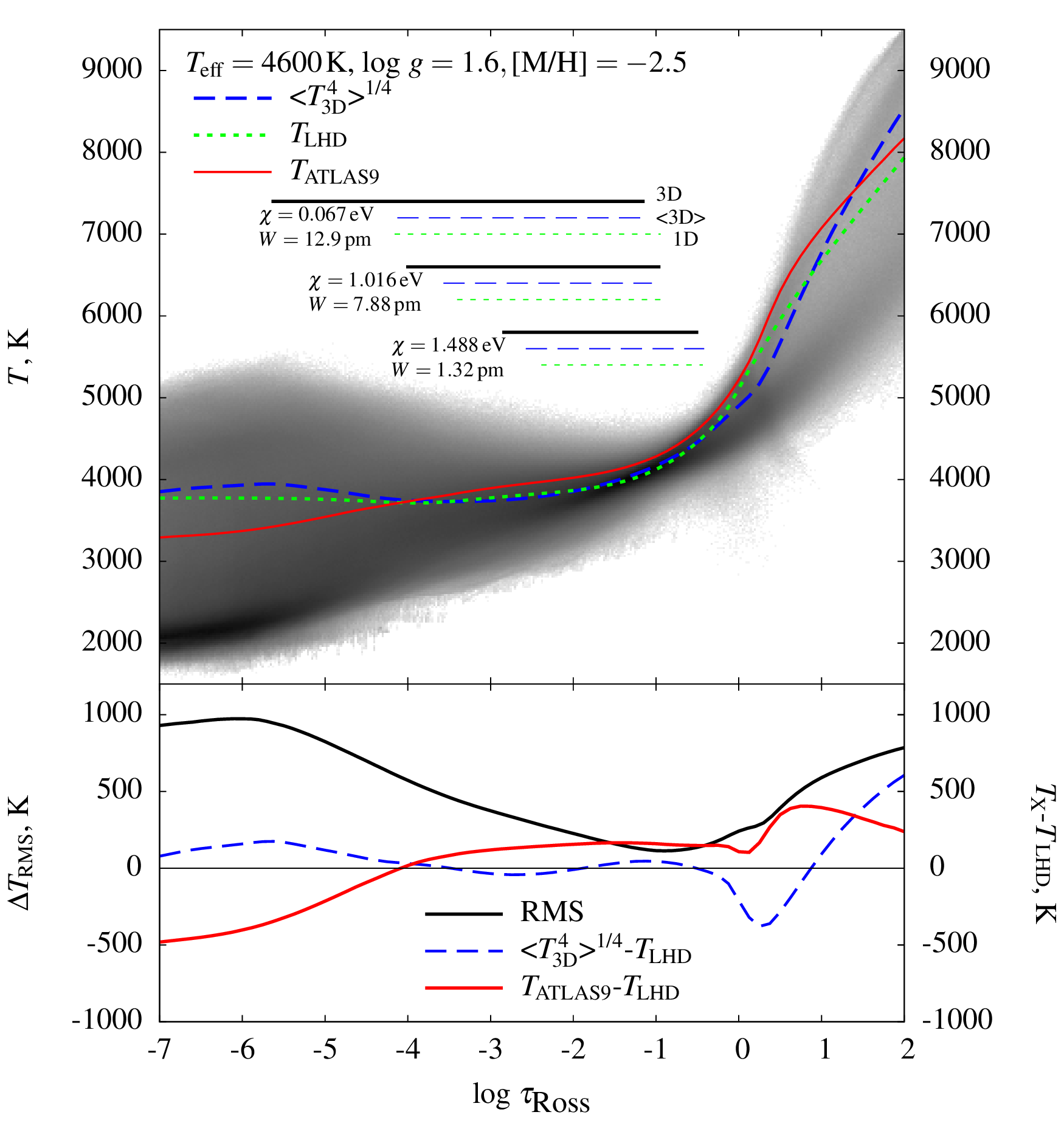}
   	\caption{\textbf{Top:} temperature structure of the 3D hydrodynamical (grey scale map of the logarithmic temperature probability density), average $\xtmean{\mbox{3D}}$ (blue dashed line), and 1D \LHD\ (green dotted line) model atmospheres of HD~122563. For comparison, we also show the temperature profile of the \ATLAS\ model atmosphere (red solid line). Horizontal bars indicate the optical depth intervals where 90\% of the equivalent width (i.e., between 5\% and 95\%) of three representative OH lines is formed in the 3D hydrodynamical (black solid bar), average $\xtmean{\mbox{3D}}$ (blue dashed bar), and 1D \LHD\ (green dotted bar) model atmospheres (spectral line parameters are given next to each set of horizontal bars). \textbf{Bottom:} RMS value of horizontal temperature fluctuations in the 3D model (black solid line, $\Delta T_{\rm RMS}$), temperature difference between the average $\xtmean{\mbox{3D}}$ and 1D models, $\xtmean{T^{\rm ~4}_{\rm 3D}} ^{\rm 1/4} - T_{\rm LHD}$ (blue dashed line), and temperature difference between the \ATLAS\ and \LHD\ model atmospheres (red solid line).}
   	\label{fig:modelat}
   \end{figure}
   
   We stress that \ATLAS\ model atmospheres are better suited for 1D~LTE abundance analysis than \LHD\ models, because of the more realistic radiative transfer scheme based on opacity distribution functions. On the other hand, \LHD\ models are better suited for assessing the influence of convection on the spectral line formation and by computing 3D--1D abundance corrections, because \LHD\ and \COBOLD\ model atmospheres are computed employing identical radiative transfer schemes and microphysics.
      
   The 3D hydrodynamical \COBOLD\ simulations were performed following the procedures described in our earlier studies \citep[see, e.g.,][]{LK12, KSL13, DKS13}. The model atmosphere was computed using a ``box-in-a-star`` setup and Cartesian grid of $160\times 160\times 300$ points ($3.85\times3.85\times3.78$\,Gm) in $x\times y\times z$, respectively ($z$ denotes the vertical direction). The radiative transfer was solved under the assumption of LTE, scattering was treated as true absorption. \MARCS\ opacities were grouped into six bins following the methodology described in \citet[][]{nordlund82,LJS94,VBS04}. The model atmosphere was computed using solar-scaled chemical composition, with the $\alpha$-element abundances enhanced by $\aoFe=+0.4\,dex$ \citep[see][for details]{DKS13}. The simulation box covered the Rosseland optical depth range of $-11 \lesssim \log \tau_{\rm Ross} \lesssim 9$. Our 3D model atmosphere was more extended than the one used in \citet{DKB15} which had the outer boundary set at $\log \tau_{\rm Ross}\approx -6.5$. The larger vertical extension in our case was needed to accommodate the formation of strong OH UV lines since in the most extreme cases it was extending to $\log \tau_{\rm Ross}\approx -7$ (one of such extreme examples is shown in Fig.~\ref{fig:contf}, left panel; see Sect.~\ref{sect:oh-form} for details).
       
   The \COBOLD\ model simulation run covered a span of $\sim$\,10 convective turn-over times as measured by the Brunt-Vaias\"{a}l\"{a} timescale \citep[see][for details]{LK12}. For further analysis, we have selected a sub-sample of 20 representative snapshots (i.e., 3D model structures computed at different instants in time). The snapshots were chosen by ensuring that the average thermodynamic and hydrodynamical properties of the 20-snapshot sub-sample and those of the entire simulation run would be as similar as possible \citep[see][for details]{LK12}.
      
   As in our previous work, we also used an average $\xtmean{\mbox{3D}}$ \COBOLD\ model atmosphere. It was obtained by averaging the fourth moment of temperature on surfaces of equal Rosseland optical depth in the 20-snapshot sub-sample. This model atmosphere was used to evaluate the influence of temperature fluctuations on the formation of OH UV lines (see Sect.~\ref{sect:3D_abund_method} below).
           
   The 1D hydrostatic \LHD\ model atmosphere was computed using the \LHD\ model atmosphere package, by utilizing the same atmospheric parameters, equation of state, opacities, and chemical composition as used in the computation of the 3D hydrodynamical \COBOLD\ model atmosphere. This was dictated by the choice of metallicities at which \MARCS\ opacities used in the \COBOLD/\LHD\ simulation runs were available. This issue is discussed further in Sect.~\ref{sect:results} but, in fact, such choice had negligible implications on the results of oxygen abundance analysis. The convective energy transport in the \LHD\ model was treated using mixing-length theory, with a mixing-length parameter of $\mlp = 1.0$. 
      
   Temperature profiles of the 3D hydrodynamical, average $\xtmean{\mbox{3D}}$, and 1D model atmospheres used in this work are shown in Fig.~\ref{fig:modelat}.

\subsection{Measurements of oxygen abundance in HD~122563 from OH UV lines}\label{sect:abund}
   
   The oxygen abundance determination in HD~122563 was carried out in several steps. First, 1D~LTE oxygen abundances were determined from the individual OH UV lines in the observed spectrum of HD~122563, by using the \ATLAS\ model atmosphere and synthetic line profiles computed with the \SYNTHE\ package (Sect.~\ref{sect:1D_abund_method}). Next, we used the \LINFOR\ spectral synthesis package together with the \LHD\ and \COBOLD\ model atmospheres to compute 1D~LTE and 3D~LTE curves of growth (COGs) for each OH UV line. The COGs were used to determine 3D--1D abundance corrections, $\Delta_{\rm 3D~LTE-1D~LTE}$ (Sect.~\ref{sect:3D_abund_method}). Finally, the 3D~LTE abundances were computed by adding 3D--1D abundance corrections to the 1D~LTE oxygen abundances determined using individual OH UV lines. The details of these procedures are summarized in the sections below.

  \begin{figure*}[!t]
  	\centering
  	\includegraphics[width=17cm]{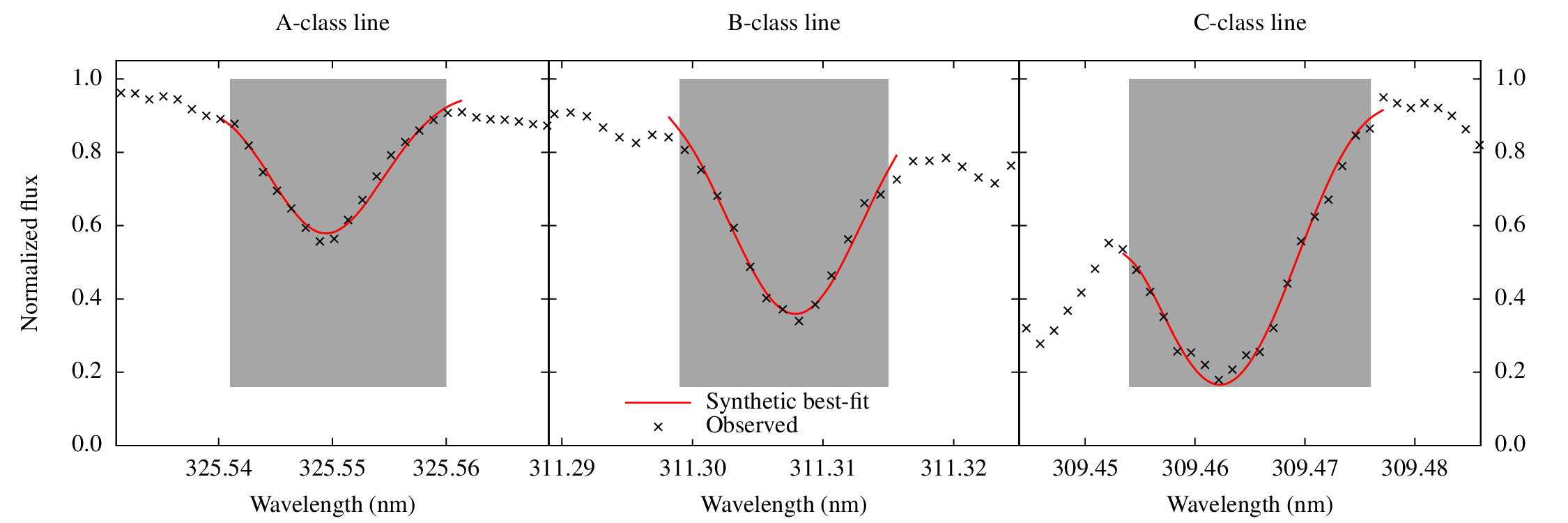}
  	\caption{Typical fits of synthetic line profiles to OH UV lines in the observed spectrum of HD~122563. The observed lines belong to different quality classes, as indicated above each panel (see Sect.~\ref{sect:1D_abund_method}for details). Black crosses show the observed spectrum, red lines are best-fitting synthetic spectra selected using a globally fixed \vbrd. Shaded areas mark the fitting regions that were selected individually for each OH UV line.}
  	\label{fig:lineclass}
  \end{figure*}

\subsubsection{Determination of 1D~LTE oxygen abundances}\label{sect:1D_abund_method}

   We determined 1D~LTE oxygen abundances by fitting theoretical line profiles to OH UV lines in the observed UVES spectrum of HD~122563.

   The list of OH UV lines used in our analysis is provided in the Table~\ref{tab:abundances}. When selecting lines for the abundance analysis we tried to make sure that they are sufficiently strong (but not saturated) and free from severe blends, so that a reliable determination of oxygen abundances was possible. Nevertheless, nearly all lines in the list are blended, some of them strongly. Therefore, we divided them into three categories according to their quality:
   \begin{list}{$\bullet$}{}
   	\item A-class lines: strong and weakly blended: 3 lines;
   	\item B-class lines: strong and moderately blended: 23 lines;
   	\item C-class lines: weak or significantly blended with lines of other chemical species: 45 lines.
   \end{list}
   
   \noindent Obviously, the most reliable oxygen abundances were determined using lines of classes A and B which constitute $\sim$35\% of all lines used in the abundance analysis. 

   Since the spectral region of interest is severely affected by strong line blends, it makes it difficult to choose an accurate continuum level in the observed spectrum of HD~122563. We therefore determined the continuum interactively, by comparing the observed spectrum with the synthetic 1D~LTE spectrum computed using the \SYNTHE\ spectral synthesis package, and seeking the best match. Obviously, the lack of accuracy of the continuum placement had an impact on the derived oxygen abundances; this issue is further discussed in Sect~\ref{sect:errors}.
   
   Theoretical profiles of OH UV lines were computed with the \SYNTHE\ package in the implementation of \citet{SBC04, Sbor05}, using the 1D \ATLAS\ model atmosphere of HD~122563 computed as described in Sect~\ref{sect:models}. The \SYNTHE\ calculations included a proper treatment of continuum scattering. For the majority of OH UV lines, atomic line parameters were taken from \citet[][]{Aoki15}. Additionally, we also used data from \citet{GGR98}, \citet{GAP06}, and Kurucz line lists (see Table~\ref{tab:abundances}). The Kurucz atomic and molecular line lists \citep{CK04} were used for computing spectral line profiles of all other chemical species. In the spectral synthesis computations we used a depth-independent microturbulence velocity of $\vmic = 2.0$\,km/s taken from \citet{SCP05}. Abundances of chemical elements other than oxygen were not determined in our analysis, despite the fact that lines of these elements were sometimes blending with OH UV lines. Instead, we used a solar-scaled chemical composition with constant $\aoFe = +0.4$ enhancement in the abundances of all $\alpha$-elements, in accordance with the results of \citet{BMS03}. All synthetic line profiles were convolved with a Gaussian profile characterized by the broadening velocity, \vbrd , which accounted for the cumulative effect of macroturbulence, stellar rotation, and instrumental broadening.
      
   Apart from the atomic parameters that were different for each OH UV line (line wavelength, oscillator strength, line excitation potential) and those parameters that were identical for all lines (e.g., microturbulence velocity), synthetic 1D~LTE profiles of individual OH UV lines were shaped by two variables -- the oxygen abundance, \logAO, and broadening velocity, \vbrd. These parameters had to be determined simultaneously, by fitting theoretical to the observed line profiles. During the fitting procedure, we also determined an arbitrary wavelength shift between the observed and synthetic spectra, ${\rm \Delta}\lambda$, in order to compensate for various wavelength-related shifts (e.g., those caused by wavelength calibration errors, imprecise spectral line wavelengths, convective wavelength shifts). All three fitting parameters -- ${\rm \Delta}\lambda$, \vbrd, and \logAO\ -- were determined by minimizing \chisq\ residuals between the observed and synthetic line profiles (see below). For this, we computed a grid of synthetic 1D~LTE OH UV line profiles which covered a range in oxygen abundances and line broadening velocities of $5.60 < \logAO < 6.80$ and $5.5 < \vbrd < 11.0$\,km/s in steps of $0.1$\,dex and $0.5$\,km/s, respectively. Tests made with finer grids have shown that the spacing of our grid was sufficient to limit the error in the oxygen abundance due to the finite step size to $\sim$0.005\,dex. This is because during the fitting procedure our grid was further interpolated to obtain better precision in the determined abundances (see below).
   
   During the fitting of the theoretical to the observed line profiles we therefore first determined ${\rm \Delta}\lambda$, \vbrd, and \logAO. This was done by comparing synthetic and observed fluxes, \Fsyn\ and \Fobs, respectively, for each individual line in our OH UV line list. The wavelength interval for the fitting of \Fobs\ and \Fsyn\ was tailored specifically in case of each OH UV line, in order to avoid blends, neighbouring lines of other elements, artefacts (see Fig.~\ref{fig:lineclass} and Sect.~\ref{sect:abund_results} below). Then, for each spectral line \textit{i} we used a number of trial values of ${\rm \Delta}\lambda^{j}$ and for each wavelength point \textit{m} in the line profile \textit{i} and each value of $\vbrd^{k}$ and $\logAO^{l}$ in our grid of synthetic line profiles we computed an array of $\chisq$ values according to
         
   \begin{equation}
      \label{eq:fitbyline}
      \chi^{2}_{i}({\rm \Delta}\lambda^{j}, \vbrd^{k}, \logAO^{l}) = \sum_{m=1}^{N^i} \frac{(\Fobs^{m} - \Fsyn^{m,j,k,l})^2}{\sigma_{\rm m}^{2}}
   \end{equation}
   
   \noindent where

   \begin{equation}
      \label{eq:sigma}
      \sigma_{m} = \frac{\Fobs^{m}}{S/N}
   \end{equation}

   \noindent is the uncertainty in the observed flux and $S/N$ is the signal-to-noise ratio in the continuum (Sect.~\ref{sect:obs}). Further, this array of \chisq\ estimates was quadratically interpolated in the $j,k,l -$\,space to obtain better precision in the determined  $\Delta\lambda^i$, $\vbrd^i$, and $\logAO^i$. The latter three values were determined individually for each OH UV line, by finding a minimum in the array of \chisq\ estimates in the $j,k,l -$\,space. The best-fitting synthetic line profiles obtained in this step were used to compute the line equivalent widths of each OH UV line, $W_i$, which were used further in the determination of 3D--1D abundance corrections and 3D~LTE oxygen abundances (Sect.~\ref{sect:3D_abund_method}). Note, however, that oxygen abundances determined in this step, $\logAO^i$, were used solely for the estimation of \Wi\ and excluded from further analysis. We also add that the individual $\Delta\lambda^i$ values are consistent with a single radial velocity within $\pm1$\,km/s.
      
   In the last step, the $\chisq_i$ estimates were summed up over all \textit{i} spectral lines to find the global best-fitting broadening velocity, $\vbrd^{\rm1D}$. We then used this fixed value of $\vbrd^{\rm1D}$, together with the individual estimates of $\Delta\lambda^i$ for each OH UV line, and recomputed $\chisq$ values again, with the only fitting parameter now being the oxygen abundance, $\logAO^i$. A search for the minimum in the array of these new $\chisq_i$ estimates yielded the final 1D~LTE oxygen abundances as derived from each oxygen line, $\logAO^{i}_{\rm 1D~LTE}$. Therefore, our final 1D~LTE oxygen abundances were determined using a fixed value of \vbrd\ with all OH UV lines. In fact, these abundances were very similar to those obtained using individual values of \vbrd, i.e., as determined in the previous step above; the largest difference between the estimates obtained using the two methods would rarely exceed $\approx0.05$\,dex and typically would be below $\approx0.01$\,dex.
   
   The final 1D~LTE oxygen abundance estimates obtained from each OH UV line, $\logAO^{i}_{\rm 1D~LTE}$, are listed in the Table~\ref{tab:abundances}. Typical fits of OH UV lines obtained during the last step in the \chisq\ minimization procedure are shown in Fig.~\ref{fig:lineclass}.
     
\subsubsection{Determination of 3D~LTE oxygen abundances using 3D--1D abundance corrections}\label{sect:3D_abund_method}

   As in \citet{DKB15}, 3D~LTE oxygen abundances were determined using 3D--1D~LTE abundance corrections, $\Delta_{\rm 3D-1D}$\footnote{The 3D--1D abundance correction, $\Delta_{\rm 3D-1D}$, is defined as difference in the oxygen abundance, \logAO, determined using the same spectral line of a given equivalent width, $W$, with the 3D hydrodynamical and 1D hydrostatic model atmospheres \citep[see, e.g.,][for details]{KSL13,DKB15}.}. To determine the abundance corrections, for each individual OH UV line, we utilized COGs that were constructed using the 3D hydrodynamical \COBOLD\ and 1D hydrostatic \LHD\ model atmospheres. Since the OH UV lines are strong (equivalent widths of the strongest lines reach $\approx 13$\,pm), they are influenced by microturbulent broadening. It was therefore critical to select a correct value of the microturbulent velocity to be used in the computations of 1D~LTE COGs with the 1D \LHD\ model atmospheres. We approached this problem in the following way.
         
\begin{figure}[!t]
   \centering
   \includegraphics[width=9cm]{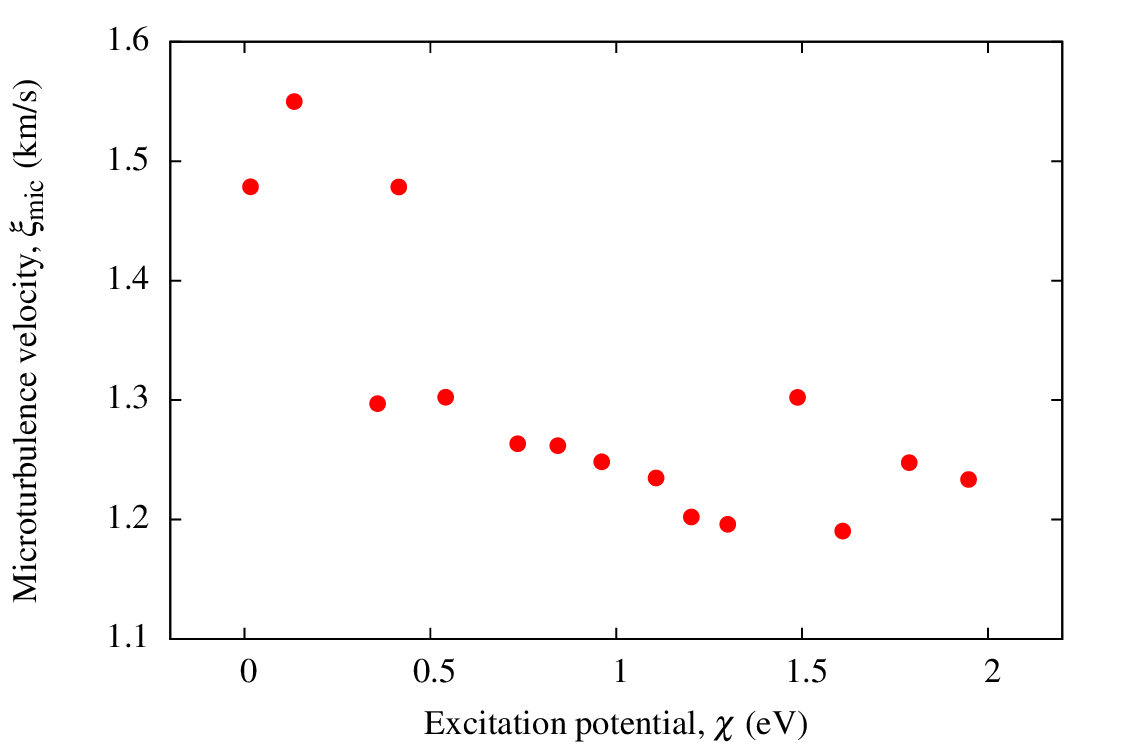}
   \caption{Microturbulence velocities derived from individual OH UV lines using the 3D hydrodynamical \COBOLD\ model atmosphere.}
   \label{fig:vmic}
   \end{figure}

   Our first step was to determine \vmic\ that would result from the 3D hydrodynamical \COBOLD\ model atmosphere of HD~122563. This was done by utilizing Method~1 from \citet{SCL13}. In short, we selected 15 OH UV lines that covered the range of excitation potentials $0 < \chi < 2$\,eV. For each of these lines, we computed 3D~LTE COGs using the \COBOLD\ model atmosphere with the original (hydrodynamic) velocity field. Subsequently, the velocity field in the \COBOLD\ model was replaced by a depth-independent and isotropic \vmic. We then utilized this modified model atmosphere to compute a number of COGs using various values of \vmic. For each OH UV line we then picked the value of \vmic\ which replicated the line equivalent width, \Wi, determined using the original \COBOLD\ model atmosphere. Fig.~\ref{fig:vmic} shows the resulting \vmic\ obtained using this approach for each of the 15 OH UV lines.
      
   It is obvious from Fig.~\ref{fig:vmic} that \vmic\ derived in this way is significantly lower than 2\,km/s that was determined in the 1D~NLTE analysis by \citet{MGS11} based on iron lines. This might be partly due to the deficiencies of the 3D model atmosphere \citep[e.g., insufficient spatial resolution, issues with artificial viscosity, and so forth; see][for a detailed discussion]{SCL13}. However, as we show in Appendix~\ref{app:vmic}, such a procedure for selecting the microturbulent velocity allows to compensate for the inadequacies in the 3D model atmosphere, in effect yielding nearly identical 3D--1D abundance corrections irrespective of \vmic\ used to obtain 1D~LTE oxygen abundances. Therefore, for the computation of 1D~\LHD\ and $\xtmean{\mbox{3D}}$ COGs -- which were further used to compute 3D--1D abundance corrections (see below) -- we used the average $\vmic = 1.30$\,km/s. The uncertainty of this estimate is $\pm0.11$\,km/s which is a standard deviation of line-to-line variation of the determined \vmic.
   
   Spectral line synthesis computations involved in the determination of 3D--1D abundance corrections were performed with the \LINFOR\ package\footnote{\url{http://www.aip.de/Members/msteffen/linfor3d}}. In this step we used the \Wi\ that were determined in the first sweep of the line profile fitting (Sect.~\ref{sect:1D_abund_method}). Abundance corrections were then obtained by measuring the difference in the oxygen abundance between the two COGs at a given value of \Wi. Finally, 3D~LTE oxygen abundances, $\logAO^{i}_{\rm 3D~LTE}$, were determined by adding the 3D--1D abundance corrections to the 1D~LTE abundances, $\logAO^{i}_{\rm 1D~LTE}$, obtained from the individual OH UV lines using \ATLAS\ model atmospheres (Sect.~\ref{sect:1D_abund_method}). The 3D--1D abundance corrections, $\Delta_{\rm 3D-1D}$, together with 3D~LTE oxygen abundances obtained from the individual OH UV lines, $\logAO^{i}_{\rm 3D~LTE}$, are provided in Table~\ref{tab:abundances}.

   It is also important to note that the carbon-to-oxygen ratio plays an important role in governing the formation of OH molecules in stellar atmospheres. When oxygen dominates over carbon, the latter is almost entirely locked in CO and only the excess of oxygen is available for the formation of other oxygen-bearing molecules, including OH \citep[see, e.g.,][]{DKS13}. According to the 1D~LTE analysis performed by \citet{SCP05}, the carbon-to-oxygen ratio in HD~122563 is ${\rm C/O} \approx 0.05$. Thus, oxygen is the majority (dominant) species in the formation of CO molecules. Even if in principle a 3D analysis may yield a C/O ratio that is slightly different from the one obtained by \citet{SCP05}, i.e., because of different 3D--1D abundance corrections for oxygen and carbon, this (supposedly, small) change in the carbon abundance should not have a noticeable effect on the strengths of the OH lines \citep[][their Appendix C]{DKS13}. Indeed, the tests that we did to verify this assumption have confirmed that changing the C/O ratio by $\pm0.05$\,dex did not alter the 3D--1D oxygen abundance corrections by more than $\sim0.01$\,dex (see Appendix~\ref{app:COratio} for test results and extended discussion)\footnote{Note, however, that the situation is entirely different when ${\rm C/O}\gtrsim 1$: in this case even small changes in the carbon abundance have a dramatic effect on the strength of OH lines. As it has been recently demonstrated by \citet{GCB16}, that in such situations the carbon and oxygen abundance should be determined in 3D simultaneously.}.

   We also note that in the \LINFOR\ calculations we treated scattering as true absorption. The effects of continuum scattering on the 3D-1D abundance corrections of OH UV lines were investigated in \citet[][see their figures 18 and 19]{HAC11}, but for hotter model atmospheres characterized by higher surface gravities. The authors found that the treatment of scattering might alter the 3D--1D abundance corrections by 0.05--0.1\,dex at the model metallicity of HD~122563. However, the authors used 1D \MARCS\ model atmospheres for reference, which does not warrant a differential comparison. Since we use \LHD\ model atmospheres for reference, we expect the effect of scattering to be smaller with respect to the results of \citet{HAC11}. A deeper investigation of the effects of continuum scattering on OH UV lines is however beyond the scope of the current work.

   For each OH UV line we also produced COGs that were constructed using the average $\xtmean{\mbox{3D}}$ model atmosphere. These COGs, together with those produced with the 3D hydrodynamical and 1D hydrostatic model atmospheres, were used to compute two additional abundance corrections, $\Delta_{\rm   3D-\langle3D\rangle}$ and $\Delta_{\rm \langle3D\rangle-1D}$. Since the $\xtmean{\mbox{3D}}$ model atmosphere is one-dimensional and is devoid of information about the horizontal inhomogeneities of thermodynamic and hydrodynamical quantities, the $\Delta_{\rm 3D-\langle3D\rangle}$ correction may be used to estimate the role of horizontal fluctuations in the line formation. Similarly, the $\Delta_{\rm \langle3D\rangle-1D}$ correction measures the role of differences between $\xtmean{\mbox{3D}}$ and 1D model atmospheres.
   
   \subsubsection{Error budget\label{sect:errors}}
   
   Four factors contribute to the uncertainty of the oxygen abundances determined from individual OH lines\begin{list}{$\bullet$}{:}

   \item formal fitting errors;
   \item errors in the atmospheric parameters;
   \item errors in the continuum placement;
   \item errors in equivalent width measurements.
   \end{list}

    \noindent In case of each individual OH UV line, the formal fitting error, $\sigma_{\rm \chisq}$, was computed by evaluating the difference in $\logAO^{i}_{\rm 1D~LTE}$ that satisfied the condition $\chisq -\chisq _{min} = 1$, with the oxygen abundances determined using a fixed value of \vbrd\ for all lines (Sect.~\ref{sect:1D_abund_method}).
   
   \begin{figure*}[!ht]
	\centering
	\includegraphics[width=14cm]{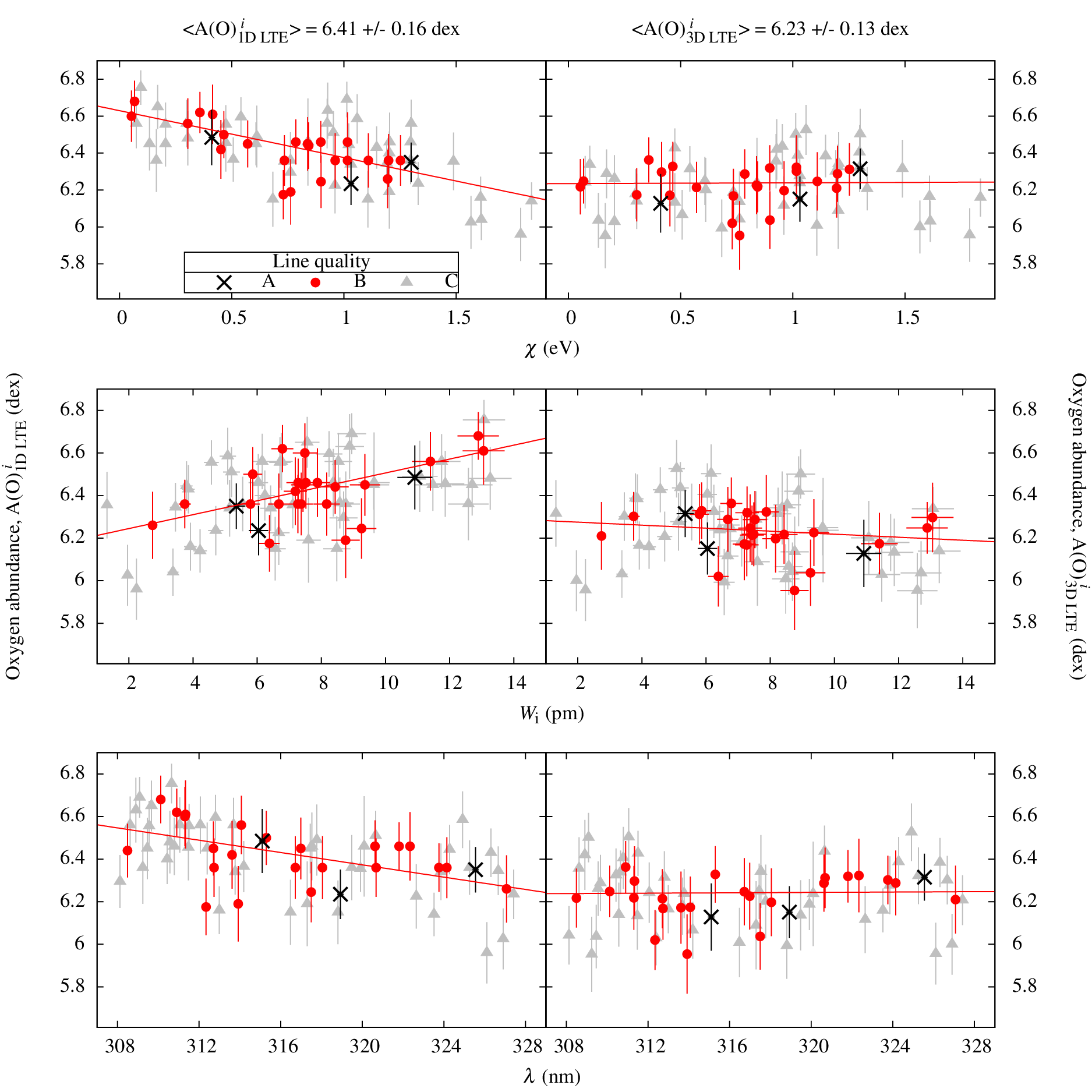}
	\caption{Oxygen abundances determined from the individual OH UV lines using a classical 1D (\ATLAS, left column) and corrected for 3D hydrodynamical effects using a \COBOLD\ model atmosphere (right column). The top, middle, and bottom panels show \logAO\ plotted versus excitation potential, $\chi$, line equivalent width, \Wi, and wavelength, $\lambda$, respectively. Different symbols mark spectral lines of different quality where A-class lines are the best/cleanest (see Sect.~\ref{sect:1D_abund_method} for details). All derived abundances are provided in Appendix~\ref{app:line_list_abund}.}
	\label{fig:abu_ABC}
\end{figure*}   
   
   \citet{CTB12} quotes errors on the determined \Teff\ and \logg\ to be 41\,K and 0.04\,dex, respectively. We therefore computed four additional \ATLAS\ model atmospheres with \Teff\ and \logg\ shifted from the values listed in Table~\ref{tab:models} by $\pm$41\,K and $\pm$0.04\,dex, respectively. The differences between the oxygen abundances obtained from each OH UV line with these model atmospheres and those determined in Sect.~\ref{sect:1D_abund_method} yielded errors due to uncertainties in the atmospheric parameters, $\sigma_{\rm \Delta Teff}$ and $\sigma_{\rm \Delta log~g}$.
 
   According to \citet{SCP05}, the error in \vmic\ is 0.2\,km/s. In order to evaluate the uncertainties in the oxygen abundances due to the imprecise determination of \vmic, we have repeated the spectral synthesis computations with \SYNTHE\ for each OH UV line, using $\vmic = 1.8$ and $2.2$\,km/s. A subsequent abundance determination using these new theoretical line profiles and comparison of the obtained abundances with those determined in Sect.~\ref{sect:1D_abund_method} provided errors due to the uncertainty in \vmic, $\sigma_{\rm vmic}^{\rm 1D}$. For most lines, this error was very small and only in rare cases exceeded 0.1\,dex.
      
   As noted in Sect.~\ref{sect:3D_abund_method}, for the calculation of $\Delta_{\rm 3D-1D}$, we used $\vmic = 1.30 \pm 0.11$\,km/s. In order to estimate the error due to the uncertainty in the determined value of \vmic , we carried out 1D~LTE line synthesis computations using the \LINFOR\ package and \LHD\ model atmosphere, again with $\vmic = 1.41$ and $1.19$\,km/s. The obtained COGs were used to determine abundance corrections $\Delta_{\rm 3D-1D}$ and, subsequently, 3D~LTE oxygen abundances. Comparison of these 3D~LTE abundances with those determined in Sect.~\ref{sect:3D_abund_method}, provided us with the error on the 3D~LTE abundances, $\sigma_{\rm vmic}^{\rm 3D}$. Again, these errors were very small and typically well below 0.03\,dex.
                                                                                                                                                                                                                                                                                                                                                           
   To estimate the error in the oxygen abundance due to the uncertainty in the continuum placement, we used a wavelength-dependent error $\sigma_{\rm flux}(\lambda)\equiv (S/N)^{-1}$ that was determined as described in  Appendix~\ref{SN-ratio}. The observed spectrum was scaled by dividing it by $1.0\pm\sigma_{\rm flux}(\lambda)$. We then repeated the 1D~LTE abundance determination with the scaled observed spectrum (following the prescription given in Sect.~\ref{sect:1D_abund_method}) and determined $\sigma_{\rm cont}$ by comparing the obtained result with that determined in Sect.~\ref{sect:1D_abund_method}.
   
   The final source of uncertainty stems from the imprecise measurements of \Wi. As discussed in Sect.~\ref{sect:3D_abund_method}, \Wi\ measurements were used solely for the estimation of the 3D--1D abundance corrections and hence they impact only $\logAO^{i}_{\rm 3D~LTE}$. We assumed that the uncertainty on \Wi\ was 5\%, a rather conservative estimate. Using this error, we recomputed the $\Delta_{\rm 3D-1D}$ abundance corrections and found that the resulting influence on $\logAO^{i}_{\rm 3D~LTE}$ never exceeded 0.025\,dex and typically was below $0.01$\,dex. Practically, this means that other sources of uncertainty dominated over the error due to uncertainty in \Wi .
      
   \begin{figure*}[!ht]
   	\centering
   	\includegraphics[width=14cm]{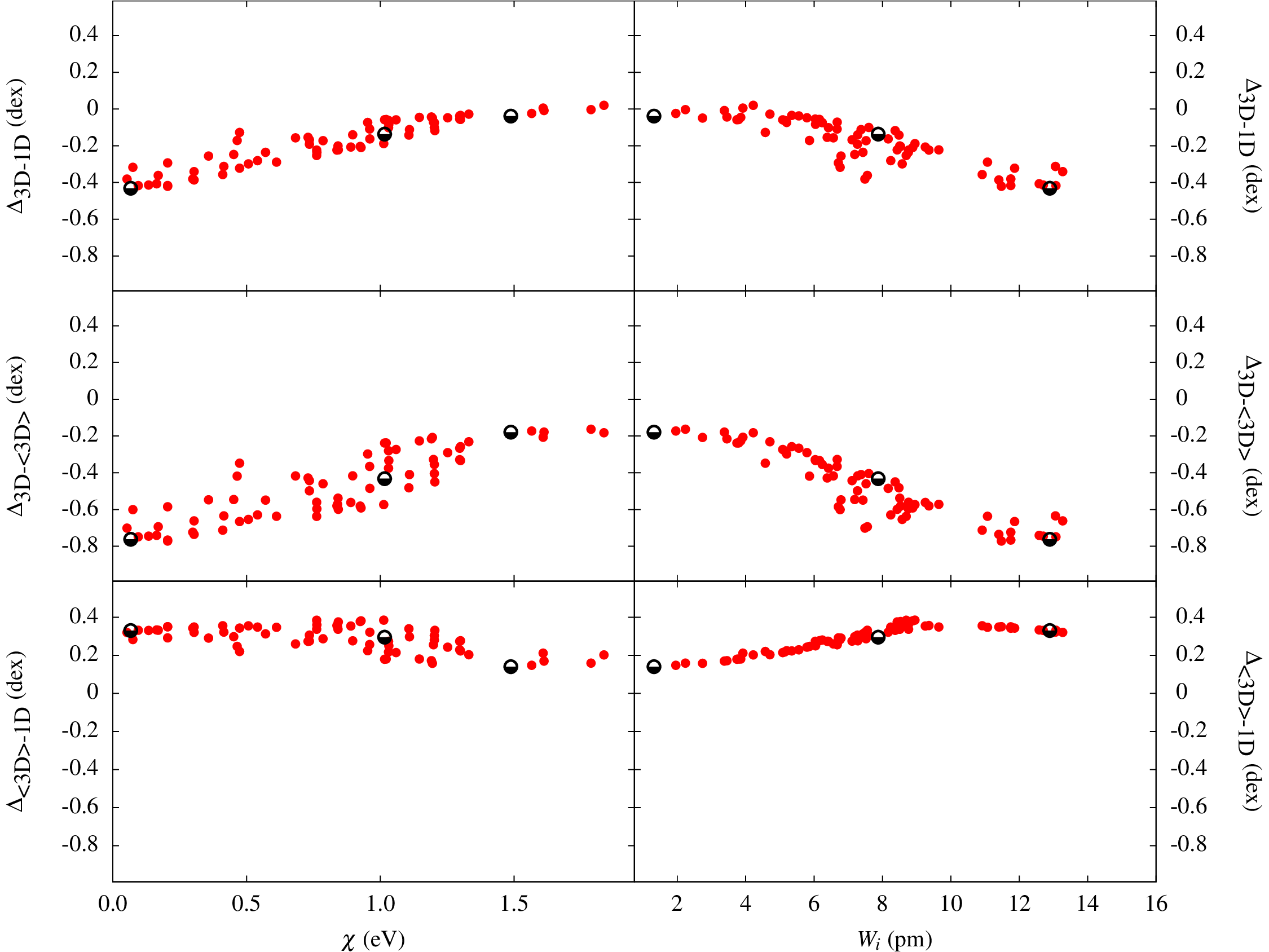}
   	\caption{The 3D--1D abundance corrections as computed for individual OH UV lines: $\Delta_{\rm 3D-1D}$ (top panels), $\Delta_{\rm 3D-\langle3D\rangle}$ (middle panels), and $\Delta_{\rm \langle3D\rangle-1D}$ (bottom panels) plotted versus the line excitation potential, $\chi$ (left), and line equivalent width, $W$ (right). Red dots mark all OH UV lines in our sample, half-filled black circles represent lines whose contribution functions are shown in Fig.~\ref{fig:contf} (see text for details). $\Delta_{\rm \langle3D\rangle-1D}$ values are provided in Appendix~\ref{app:line_list_abund}.}
   	\label{fig:dabu_main}
   \end{figure*}

   The final errors in oxygen abundances as determined from each OH UV line were computed as a square root of the sum of relevant errors in quadratures and are given in Table~\ref{tab:abundances}.
      
\begin{table}
	\caption{1D and 3D abundances of OH UV and IR and [\ion{O}{i}] lines in HD~122563. Oxygen abundances determined using IR lines are from \citet{DKB15}, while those determined using the [\ion{O}{i}] line was taken from \citet{SCP05}}\label{tab:abushort}
	\centering
	\begin{tabular}{|c|c|c|c|c|}
		\hline 
		& \multicolumn{2}{|c|}{\logAO } & \multicolumn{2}{|c|}{ \OoFe } \\
		\hline
		& 1D~LTE & 3D~LTE & 1D~LTE & 3D~LTE \\ 
		\hline 
		OH UV        & 6.41 $\pm$ 0.16 & 6.23 $\pm$ 0.13 & 0.25 $\pm$ 0.16 & 0.07 $\pm$ 0.13 \\ 
		\hline 
		OH IR        & 6.63 $\pm$ 0.10 & 6.39 $\pm$ 0.11 & 0.47 $\pm$ 0.10 & 0.23 $\pm$ 0.11 \\ 
		\hline 
		[\ion{O}{i}] & 6.54 $\pm$ 0.15 & 6.53 $\pm$ 0.15  & 0.38 $\pm$ 0.15 & 0.37 $\pm$ 0.15 \\ 
		\hline 
	\end{tabular} 
\end{table}

   \begin{figure*}[!t]
   	\centering
   	\includegraphics[width=17cm]{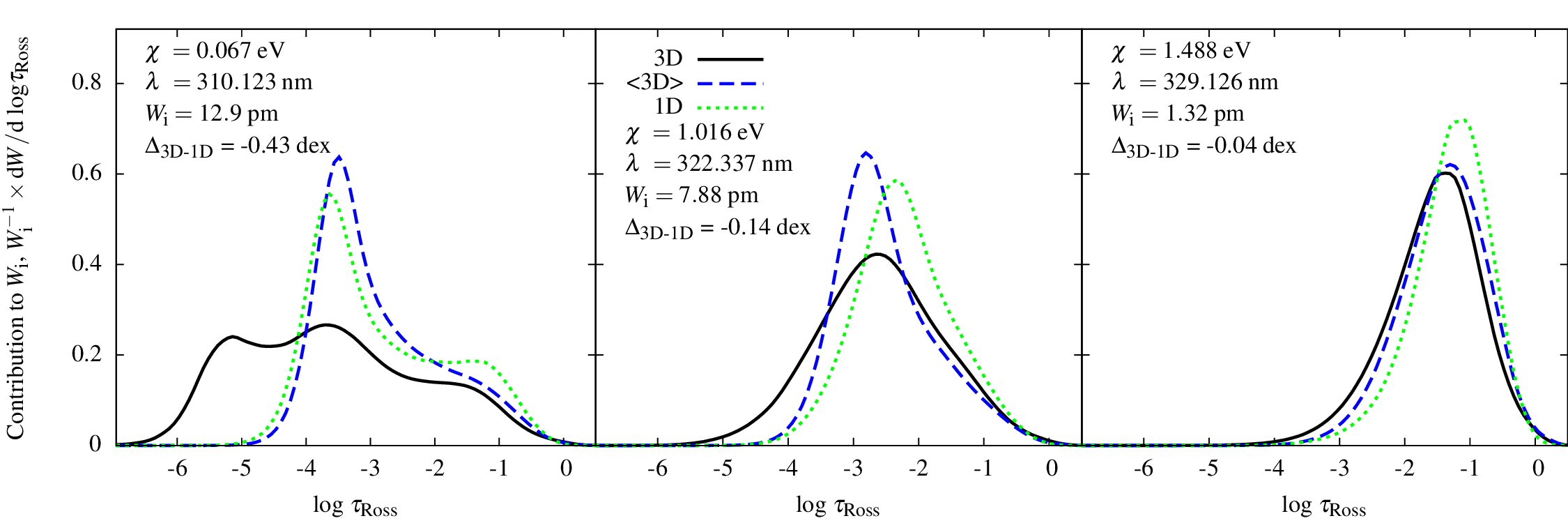}
   	\caption{Normalized 3D, $\langle3D\rangle$ and 1D contribution functions to equivalent width for the three OH UV lines marked in Fig.~\ref{fig:dabu_main}. The parameters of these spectral lines sample the entire range of $\chi$ and \Wi\ values of OH UV lines used in our work. See Table~\ref{tab:abucorr} and text for details.}
   	\label{fig:contf}
   \end{figure*}

%-------------------------------------------------------------------------
\section{Results and discussion}\label{sect:results}

\subsection{1D~LTE and 3D~LTE oxygen abundance in HD~122563}\label{sect:abund_results}

The 1D~LTE oxygen abundances obtained from individual OH UV lines are plotted against the line excitation potential, equivalent width, and central wavelength in Fig.~\ref{fig:abu_ABC} (left panels). The weighted least squares fitting of 1D~LTE abundances yielded a slope $d\logAO_{\rm 1D~LTE}/d\chi = -0.25 \pm 0.03$ dex/eV. This result is very similar to the one derived by \citet{Aoki15}, who has analysed the extremely metal-poor subgiant star BD+44$^{\circ}$493 and derived $d\logAO_{\rm 1D~LTE}/d\chi = -0.23 \pm 0.03$ dex/eV. Applying the Student's $t$-test to our data, the null hypothesis that the slope between abundance and $\chi$ is zero can be rejected at {\it a posteriori} significance level $\alpha=0.01$\%.

These results demonstrate that 1D~LTE oxygen abundance depends on these line parameters indicating systematic errors in the modelling. For example, the negative slope in the $\logAO^{i}_{\rm 1D~LTE} - \chi$ plane may indicate that the model atmosphere is too hot, whereas accurate flux measurements \citep{CTB12} and non-LTE analysis of iron lines \citep{MGS11} suggest that this is not the case. The lowest quality lines (class C) typically show a larger line-to-line scatter of the oxygen abundances. The trends, however, are seen in cases of both best (classes A--B) and lower (class C) quality lines. Similar results have been obtained earlier by other authors who also noticed that the oxygen abundance determined from OH lines either in the UV or IR showed trends with the line parameters, such as excitation potential and/or line equivalent width \citep[e.g.,][]{MB02,Aoki15,DKB15}\footnote{Trends seen in the different panels of Fig.~\ref{fig:abu_ABC} are not entirely independent. For example, lines with the highest excitation potential are also the weakest, thus negative slope seen in the $\logAO^{i}_{\rm 1D~LTE} - \chi$ plane translates into the opposite trend seen in the $\logAO^{i}_{\rm 1D~LTE} - W_{\rm i}$ plane.}.

The situation looks significantly better when 3D~LTE oxygen abundances are used instead (Fig.~\ref{fig:abu_ABC}, right panels). The trends now virtually vanish, with $d\logAO_{\rm 3D~LTE}/d\chi = 0.00 \pm 0.04$ dex/eV. This is confirmed by the Student's $t$-test which yields a {\it a posteriori} significance level $\alpha=15$\% computed using our data. This large number gives a clear indication that the rejection of the null hypothesis is not warranted. In this context it is interesting to note that \citet{DKB15} have also applied 3D--1D abundance corrections to their 1D~LTE oxygen abundances determined from OH IR lines in four-metal poor giants, including HD~122563. In their case, however, this procedure did not help to entirely remove the trends seen in the $\logAO^{i}_{\rm 1D~LTE} - \chi $ plane, although one may also argue that the number of OH IR lines used in their study was small and thus it was difficult to judge whether these trends were statistically significant in the first place.

For comparison, we collected 1D~LTE and 3D~LTE oxygen abundances determined in HD~122563 using OH UV/IR lines \citep[this work and][]{DKB15} and the forbidden [\ion{O}{i}] line (literature data) in Table~\ref{tab:abushort} (in case of OH UV/IR lines we provide abundances that are weighted averages of measurements from individual lines; the RMS error given in Table~\ref{tab:abushort} measures the spread in individual abundances). One may conclude that there is a fair agreement between the 3D~LTE oxygen abundances obtained using OH UV and IR lines. On the other hand, there is a noticeable difference between the 3D~LTE oxygen abundance determined from OH UV lines and that obtained from the forbidden [\ion{O}{i}] line. It is possible that these differences may point to the importance of non-LTE effects in the formation of OH lines which were not properly taken into account in our study of OH UV/IR lines. Indeed, OH UV/IR lines form in the outer atmosphere where, for example, overdissociation due to the non-local UV radiation field may lead to smaller number densities of OH molecules (i.e., with respect to those expected in LTE), and thus, to slightly weaker lines in 3D~NLTE than those in 3D~LTE \citep[see discussion in][]{AG01}. This would result in less negative 3D--1D abundance corrections and, consequently, to higher oxygen abundances obtained from OH lines -- which would bring them into better agreement with those obtained from the forbidden [\ion{O}{i}] line. Moreover, the 3D non-LTE effects may also influence the slope in the $\logAO^{i}_{\rm 3D~LTE}$ vs. $\chi$ plane, as the size of these effects may be different for lines forming at different optical depths in the atmosphere \citep[see, e.g., the 3D non-LTE study of iron line formation by][]{ALA16}. However, a detailed investigation of this problem is beyond the scope of the present study.
     
   \begin{table}[!t]
   	\caption{Line parameters, equivalent widths and abundance corrections -- $\Delta_{\rm 3D-1D}$, $\Delta_{\rm 3D-\langle3D\rangle}$ $\Delta_{\rm \langle3D\rangle-1D}$ -- of the OH UV lines shown in Fig.~\ref{fig:contf}.}\label{tab:abucorr}
   	\centering
   	\begin{tabular}{|c|c|c|c|c|c|}
   		\hline 
   		$\lambda$ & $\chi$ & \Wi\ & $\Delta_{\rm 3D-1D}$ & $\Delta_{\rm 3D-\langle3D\rangle}$ & $\Delta_{\rm \langle3D\rangle-1D}$\\ 
   		nm        & eV     & pm          & dex & dex & dex \\ 
   		\hline 
   		310.123 & 0.067 & 12.9 & -0.43   & -0.76  & 0.33 \\ 
   		322.337 & 1.016 & 7.88 & -0.14   & -0.43  & 0.29 \\ 
   		329.126 & 1.488 & 1.32 & -0.04   & -0.18  & 0.14 \\ 
   		\hline 
   	\end{tabular} 
   \end{table}
   
\subsection{OH UV line formation \label{sect:oh-form}}

   To better understand the OH UV line formation in the 3D hydrodynamical and 1D hydrostatic model atmospheres we focus on the abundance corrections $\Delta_{\rm 3D-1D}$ and their components -- $\Delta_{\rm 3D-\langle3D\rangle}$ and $\Delta_{\rm \langle3D\rangle-1D}$ (Fig.~\ref{fig:dabu_main}). It is evident that the $\Delta_{\rm 3D-1D}$ correction depends on both $\chi$ and \Wi . This is in contrast with the findings of \citet[][]{DKB15} for OH IR lines where no such dependence has been found. In our case, the $\Delta_{\rm 3D-1D}$ corrections are negative for low excitation lines but then progressively approach 0.0\,dex for higher excitation lines. The plot also indicates that the $\Delta_{\rm 3D-1D}$ corrections tend to decrease with \Wi . Such behaviour is not unexpected since in our OH UV line sample line strength anti-correlates with the line excitation potential, i.e., lines with lower $\chi$ values are stronger (see Appendix~\ref{app:abucorr_add}). We have not found any significant relation between the abundance corrections and line wavelength though. This, again, is plausible as the wavelength region where the OH UV lines are located is rather narrow and thus the continuum opacity (as well as line formation depth) is similar for all OH UV lines.
   
   The trends seen in Fig.~\ref{fig:dabu_main} are caused by an interplay between several factors (see Appendix~\ref{app:abucorr_add} for details): 
   \begin{itemize}
   \item
   lines with the lowest excitation potentials are on average strongest. Their formation therefore extends farthest into the outer atmosphere where temperature fluctuations are largest (cf. Fig.~\ref{fig:contf} and \ref{fig:modelat}, where in Fig.~\ref{fig:contf} we show contribution functions\footnote{Contribution function of a given spectral line is defined here as $d W/d \log \tau_{\rm Ross}$ and thus measures the rate at which line equivalent width grows at any given optical depth \citep[see, e.g.,][]{Magain86}} of three different OH UV lines characterized by different $\chi$ and \Wi ; line parameters are summarized in Table~\ref{tab:abucorr}). This leads to stronger lines in 3D and, thus, to largest and negative $\Delta_{\rm 3D-\langle3D\rangle}$ abundance corrections. Since with increasing $\chi$ (and thus decreasing \Wi ) the outer boundary of the line formation region slowly shifts to higher optical depths, the influence of temperature fluctuations becomes less important there and thus $\Delta_{\rm 3D-\langle3D\rangle}$ corrections gradually decrease in their magnitude;
   \item   
on the other hand, while the difference between the average $\xtmean{\mbox{3D}}$ and 1D temperature profiles is small and varies little over the line formation region, there is a systematic difference in the temperature gradient in the continuum forming layers in the two model atmospheres (around $\log \tau_{\rm Ross}\approx 0$, see Fig.~\ref{fig:modelat}). This not only leads to differences in the radiative continuum flux in the UV produced by the $\xtmean{\mbox{3D}}$ and 1D model atmosphere, respectively, but also translates into different slopes of the linear parts of the respective COGs. Obviously, the abundance corrections derived from two curves of growth characterised by different slopes must be proportional to the line strength, thus explaining the very tight correlations between the $\Delta_{\rm \langle3D\rangle-1D}$ correction and \Wi .
   \end{itemize}
   
      One may therefore conclude that the complex dependence of the total abundance correction, $\Delta_{\rm 3D-1D}$, on both $\chi$ and \Wi\ is mostly due to the fact that OH UV lines used in our study cover a wide range both in $\chi$ and \Wi . This also explains why the $\Delta_{\rm 3D-1D}$ abundance corrections were largely independent of $\chi$ in \citet[][]{DKB15}. All OH IR used in the latter study were very weak and thus formed at similar optical depths which resulted in similar abundance corrections for all lines.
 
\subsection{Early galactic evolution of oxygen: 3D~LTE oxygen abundances from OH UV/IR lines}

   \begin{figure}[!t]
   \centering
   \includegraphics[width=9cm]{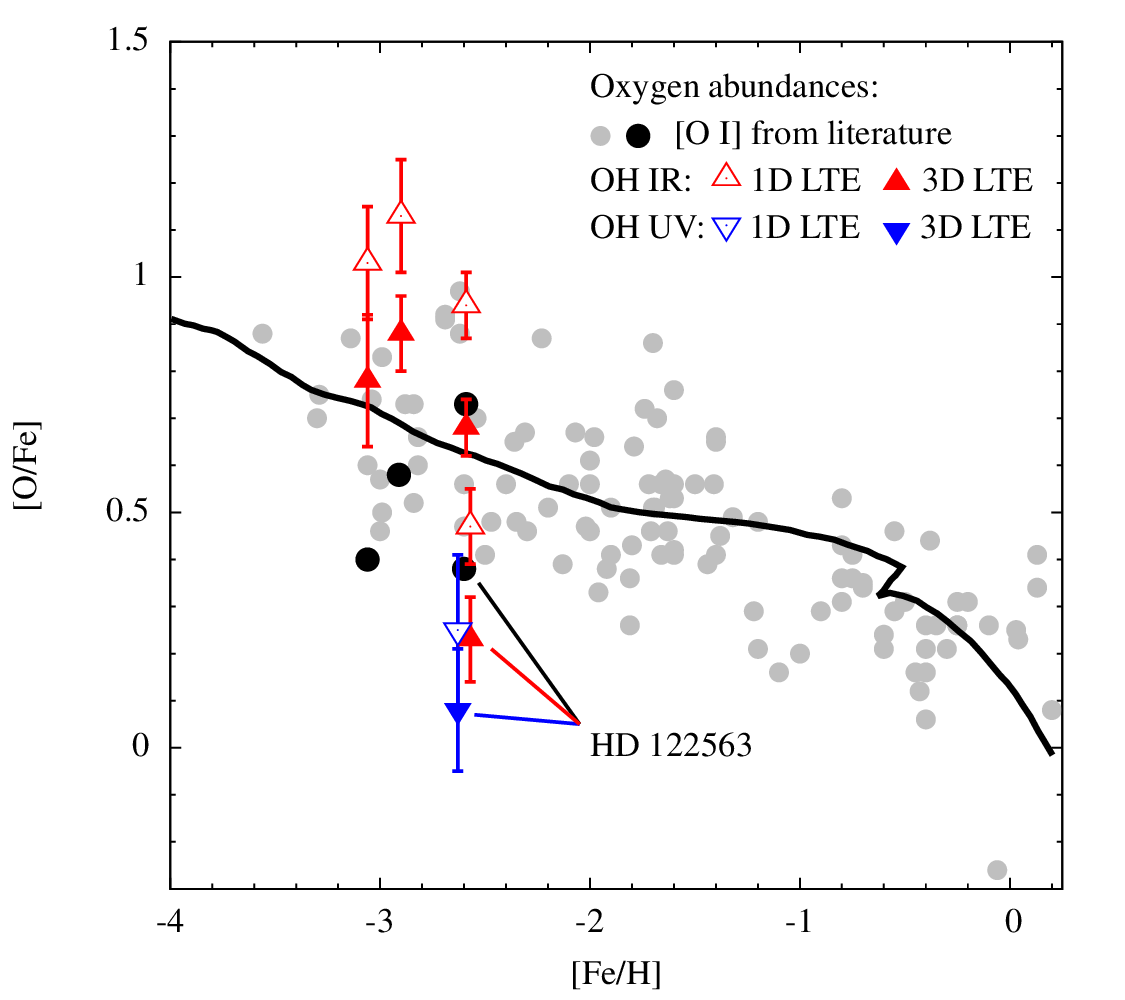}
   \caption{Oxygen-to-iron ratios in the Galactic metal-poor stars (following Fig.~4 from \citealt{DKB15}). Solid grey circles depict $\OoFe$ ratios in red giants, subgiants, and main sequence stars determined using the forbidden \ensuremath{\left[\ion{O}{i}\right]} 630~nm line \citep[literature data; see][for details]{DKB15}. Red triangles show 3D~LTE and 1D~LTE $\OoFe$ ratios in four metal-poor giants obtained from OH IR lines by \citet[][open and filled symbols, respectively]{DKB15}. The $\OoFe$ ratios determined in these stars from \ensuremath{\left[\ion{O}{i}\right]} line are marked as black solid circles. Blue triangles are average $\OoFe$ ratios in HD~122563 obtained in this study from 71 OH UV lines (for readability, abundances determined from OH UV and IR lines are shifted horizontally by $\pm$0.03\,dex). The solid line is the evolutionary model of \citet{FMC04}. Note that all abundance ratios are given in the reference scale where the solar oxygen abundance is $A({\rm O}) = 8.76$ \citep[][]{SPC15}.}   
   \label{fig:galev}
   \end{figure}

   The mean \OoFe\ ratios determined in HD~122563 using OH UV/IR lines and the forbidden [\ion{O}{i}] line are summarized in Table~\ref{tab:abushort}. To compute these \OoFe\ ratios we used the 1D~NLTE estimate of $\FeoH = -2.60$ from \citet{MGS11} and the recently determined 3D~NLTE solar oxygen abundance of $\logAO_{\odot} = 8.76$ from \citet{SPC15}\footnote{If the 3D~NLTE Solar oxygen abundance of $\logAO_{\odot} = 8.66$ determined by \citet{A05} were used instead, the \OoFe\ ratios listed in Table~\ref{tab:abushort} would increase by $+0.1$\,dex}. In the context of our study, it would be preferable to account for the 3D and non-LTE effects on the estimate of \FeoH, too. However, this would require a full 3D~NLTE study of iron abundance because, as shown by \citet{KKS16}, simply adding 3D--1D abundance corrections to the 1D~NLTE abundance estimates may lead to inconsistent results. Such analysis is certainly beyond the scope of the present study.
   
   \OoFe\ values determined in HD~122563 using OH UV (this work) and OH IR lines \citep{DKB15} suggest a mean (unweighted) value of $\OoFe = 0.16\pm0.12$. Oxygen abundance estimate obtained by \citet{SCP05} using the forbidden [\ion{O}{i}] line yields $\OoFe = 0.37\pm0.15$. Although, the oxygen-to-iron abundance ratios obtained from UV and IR lines are somewhat different, none of the values support the high over-abundance of oxygen in the early Galaxy, as argued by, e.g., \citet{IRG01}. On the other hand, the two values are noticeably lower than the typical oxygen-to-iron abundance ratio that would be expected for the Galactic field stars at this metallicity, as seen in Fig.~\ref{fig:galev}. This picture is also supported by the results of \citet{JGH10} and \citet{AAC15} who used 3D hydrodynamical model atmospheres to derive 3D~LTE/NLTE oxygen abundances in a sample of Galactic subgiant and/or dwarf stars; the latter authors determined $\OoFe \approx 0.4-0.6$ at the metallicity of HD~122563. Obviously, it would be premature to draw any firm conclusions based on the analysis of a single star. Nevertheless, it seems that the oxygen abundance in HD~122563 loosely fits the scenario proposed by \citet{FMC04}, albeit on the lower boundary of the oxygen-to-iron ratios observed at this metallicity.

%-------------------------------------------------------------------------
\section{Conclusions}\label{sect:conclusions}

    In this study we determined the abundance of oxygen in the atmosphere of the metal-poor late-type giant HD~122563 using 71 OH UV lines located in the wavelength range of 308--330\,nm. Our analysis was carried out by fitting synthetic 1D~LTE line profiles computed using a \ATLAS\ model atmosphere and the \SYNTHE\ spectral synthesis package to the OH UV line profiles in the observed high-resolution UVES spectrum of HD~122563. The fitting procedure was done using a \chisq\ minimization technique. We then utilized 3D hydrodynamical \COBOLD\ and 1D hydrostatic \LHD\ model atmospheres in order to compute 3D--1D abundance corrections which were subsequently used to determine 3D~LTE oxygen abundances. The average 3D~LTE oxygen abundance obtained in HD~122563 from 71 OH UV lines is $A{\rm (O)_{\rm 3D~LTE}}=6.23\pm0.13$. Based on this value, we further computed an oxygen-to-iron ratio for which we used $\FeoH = -2.60$ from \citet{MGS11} and a Solar oxygen abundance of $\logAO_{\odot} = 8.76$ determined by \citet{SPC15} using a \COBOLD\ model atmosphere and 3D~NLTE spectrum synthesis techniques. This yielded $\OoFe=0.07\pm0.13$ as determined from OH UV lines in HD~122563.
   
    Our 1D~LTE analysis has revealed that oxygen abundances obtained from OH UV lines were clearly dependent on the spectral line parameters, such as excitation potential, $\chi$, and line equivalent width, $W$. Additionally, 1D~LTE abundances determined from OH UV and OH IR lines failed to agree and showed differences of up to $\approx 0.3$\,dex. Besides, the two abundance estimates did not agree with the one determined using the forbidden [\ion{O}{i}] line which is known to be insensitive to 3D hydrodynamical and NLTE effects \citep[see, however, ][]{DAK16}. These results are in line with the findings of other authors who have encountered similar problems in 1D~LTE oxygen abundance analyses based on OH UV/IR lines \citep[e.g.,][]{BMS03,Aoki15}.
   
    This situation has improved significantly when 1D~LTE abundances were corrected for the 3D-related effects which was done by applying 3D--1D abundance corrections. Such procedure has helped to minimize significantly the trends of the oxygen abundance with the spectral line parameters as seen in the results of 1D analysis. This indicates that 3D hydrodynamical effects are indeed important in the formation of OH UV lines in the metal-poor red giant atmospheres and therefore should be properly taken into account in oxygen abundance analysis based on OH UV/IR lines. On the other hand, 3D~LTE oxygen abundance obtained in HD~122563 using OH UV lines, $\OoFe = 0.07\pm0.13$, was found to be lower than that determined using the forbidden [\ion{O}{i}] line, $\OoFe = 0.37\pm0.15$ (although both values are consistent at the $1\sigma$ level). While the exact cause for this discrepancy remains unclear, it may be at least partly caused by the non-LTE effects on OH lines which are known to be important in the outer atmospheres of the metal-poor stars, such as HD~122563, but which were not taken into account in the present study.
   
    The mean (unweighted) oxygen-to-iron ratio determined in HD~122563 using OH UV (this study) and OH IR lines \citep{DKB15}, $\OoFe = 0.16\pm0.12$, as well as one estimate obtained by \citet{SCP05} using the forbidden [\ion{O}{i}] line, $\OoFe = 0.37\pm0.15$, do not seem to support the high over-abundance of oxygen in the metal-poor stars. In fact, it seems that the oxygen abundance in HD~122563 falls on the lower boundary of the oxygen-to-iron ratio distribution observed in Galactic field stars at this metallicity. Our results obtained from the study of OH UV/IR lines therefore suggest that oxygen overabundance higher than $\OoFe \approx 1.0$ seems to be unlikely in the metal poor giants at $-3.0 < \FeoH < -2.5$. It is clear, however, that analyses of larger samples of metal-poor giants carried out with 3D hydrodynamical model atmospheres would be very desirable in order to acquire a better understanding of the evolution of oxygen in the early Galaxy.
   	 
%__________________________________________________________________

\begin{acknowledgements}
{This work was supported by grant from the Research Council of Lithuania (MIP-089/2015). HGL acknowledges financial support by Sonderforschungsbereich SFB 881 "The Milky Way System" (subproject A4) of the German Research Foundation (DFG). A part of computations were performed at the High Performance Computing Center, HPC Sauletekis, of the Faculty of Physics, Vilnius University. All plots in this paper were prepared using free Gnuplot software. Results presented in this paper are based on observations collected at the European Organisation for Astronomical Research in the Southern Hemisphere under ESO programme 266.D-5655. We also thank the anonymous referee for the comments that have helped to improve the paper.}
\end{acknowledgements}

%__________________________________________________________________

\bibliographystyle{aa}

\begin{appendix}

   \section{OH line list and oxygen abundances determined from individual lines}\label{app:line_list_abund}

   In this Section we provide a list of OH UV lines that were used in the oxygen abundance determination in HD~122563. We also provide oxygen abundances obtained from individual OH UV lines, as well as their errors. All this information is summarized in Table~\ref{tab:abundances} the contents of which are as follows:

   \begin{itemize} 
   \item column~1: line central wavelength;
   \item column~2: line excitation potential;
   \item column~3: log~$gf$ value;
   \item column~4: source of log~$gf$ value;
   \item column~5: quality of the line;
   \item column~6: equivalent width;
   \item column~7: 1D~LTE oxygen abundance and its error;
   \item column~8: 3D--1D abundance correction and its error;
   \item column~9: 3D~LTE oxygen abundance and its error.
   \end{itemize}

\onecolumn
{\centering

\tablehead{
\hline
$\lambda$ & $\chi$ & log~$gf$ & Source  & Quality & \Wi\  & \logAOOD & $\Delta_{\rm 3D-1D}$ & \logAOTD  \\ 
    nm    & eV     &          &         &         & pm    & dex       & dex                  & dex \\ 
\hline
}

\tabletail{
\hline
\multicolumn{9}{|c|}{\small\sl continued on next page}\\
\hline
}

\tablelasttail{
\hline
\multicolumn{9}{|l|}{Note: 'source' column indicates the reference of log \textit{gf} value.} \\
\multicolumn{9}{|l|}{WA - \citet{Aoki15}, RK - \citet{CK04}, GI - \citet{GGR98}, GP - \citet{GAP06}}\\
\hline
}
\tablecaption[]{OH UV line parameters and 1D/3D abundances derived from different spectral lines with 1D and 3D model atmospheres.}
\begin{supertabular}{|c|c|c|c|c|c|c|c|c|}
\label{tab:abundances}   
308.125 & 0.762 & -1.882 & WA & C & 8.69   & 6.30$\pm$0.12 &  -0.26$\pm$0.05 & 6.04$\pm$0.13 \\
308.489 & 0.843 & -1.874 & WA & B & 8.43   & 6.44$\pm$0.12 &  -0.22$\pm$0.05 & 6.22$\pm$0.13 \\
308.622 & 0.925 & -1.852 & WA & C & 8.52   & 6.56$\pm$0.13 &  -0.20$\pm$0.05 & 6.36$\pm$0.14 \\
308.900 & 0.928 & -1.869 & WA & C & 8.88   & 6.63$\pm$0.15 &  -0.21$\pm$0.06 & 6.42$\pm$0.16 \\
309.086 & 1.013 & -1.850 & WA & C & 8.94   & 6.69$\pm$0.09 &  -0.19$\pm$0.06 & 6.50$\pm$0.11 \\
309.239 & 0.164 & -1.782 & WA & C & 12.5   & 6.36$\pm$0.17 &  -0.41$\pm$0.02 & 5.95$\pm$0.17 \\
309.461 & 0.134 & -1.899 & WA & C & 12.7   & 6.45$\pm$0.14 &  -0.41$\pm$0.02 & 6.04$\pm$0.14 \\
309.554 & 0.205 & -3.121 & WA & C & 6.71   & 6.56$\pm$0.11 &  -0.29$\pm$0.04 & 6.26$\pm$0.12 \\
309.665 & 0.170 & -3.098 & WA & C & 7.55   & 6.65$\pm$0.11 &  -0.36$\pm$0.05 & 6.29$\pm$0.13 \\
310.123 & 0.067 & -2.205 & WA & B & 12.8   & 6.68$\pm$0.11 &  -0.43$\pm$0.04 & 6.25$\pm$0.12 \\
310.435 & 1.202 & -1.868 & WA & C & 6.24   & 6.40$\pm$0.11 &  -0.07$\pm$0.04 & 6.33$\pm$0.12 \\
310.566 & 0.304 & -1.708 & WA & C & 13.2   & 6.48$\pm$0.14 &  -0.34$\pm$0.02 & 6.14$\pm$0.15 \\
310.654 & 0.095 & -2.142 & WA & C & 13.0   & 6.76$\pm$0.09 &  -0.42$\pm$0.04 & 6.34$\pm$0.10 \\
310.785 & 1.297 & -1.858 & WA & C & 6.03   & 6.46$\pm$0.13 &  -0.05$\pm$0.04 & 6.41$\pm$0.14 \\
310.907 & 0.358 & -3.129 & RK & B & 6.78   & 6.62$\pm$0.11 &  -0.26$\pm$0.04 & 6.36$\pm$0.12 \\
311.053 & 1.300 & -1.873 & WA & C & 6.15   & 6.56$\pm$0.13 &  -0.06$\pm$0.04 & 6.50$\pm$0.13 \\
311.307 & 0.053 & -3.274 & RK & B & 7.48   & 6.60$\pm$0.13 &  -0.38$\pm$0.05 & 6.22$\pm$0.15 \\
311.336 & 0.415 & -1.649 & WA & B & 13.0   & 6.61$\pm$0.16 &  -0.31$\pm$0.02 & 6.30$\pm$0.16 \\
311.477 & 0.474 & -1.594 & WA & C & 11.8   & 6.46$\pm$0.15 &  -0.32$\pm$0.03 & 6.13$\pm$0.15 \\
311.507 & 0.474 & -3.170 & RK & C & 4.57   & 6.56$\pm$0.10 &  -0.13$\pm$0.02 & 6.43$\pm$0.10 \\
312.058 & 0.075 & -3.280 & RK & C & 6.76   & 6.56$\pm$0.13 &  -0.32$\pm$0.04 & 6.24$\pm$0.14 \\
312.343 & 0.730 & -2.102 & RK & C & 6.38   & 6.18$\pm$0.13 &  -0.15$\pm$0.04 & 6.02$\pm$0.14 \\
312.394 & 0.205 & -2.003 & WA & C & 11.4   & 6.45$\pm$0.12 &  -0.42$\pm$0.03 & 6.03$\pm$0.12 \\
312.704 & 0.571 & -2.442 & WA & B & 7.42   & 6.45$\pm$0.12 &  -0.24$\pm$0.05 & 6.21$\pm$0.13 \\
312.735 & 0.735 & -2.269 & WA & B & 7.26   & 6.36$\pm$0.13 &  -0.19$\pm$0.05 & 6.17$\pm$0.14 \\                                                                     
312.806 & 0.541 & -2.511 & WA & C & 8.24   & 6.60$\pm$0.10 &  -0.28$\pm$0.05 & 6.31$\pm$0.12 \\
312.993 & 1.609 & -1.490 & RK & C & 3.91   & 6.16$\pm$0.11 &   0.01$\pm$0.01 & 6.16$\pm$0.11 \\
313.617 & 0.452 & -2.570 & WA & C & 7.18   & 6.42$\pm$0.15 &  -0.25$\pm$0.05 & 6.17$\pm$0.16 \\
313.689 & 0.299 & -1.939 & WA & C & 11.7   & 6.56$\pm$0.12 &  -0.38$\pm$0.03 & 6.18$\pm$0.13 \\
313.770 & 1.031 & -1.954 & RK & C & 6.42   & 6.34$\pm$0.10 &  -0.10$\pm$0.04 & 6.24$\pm$0.11 \\
313.916 & 0.763 & -1.564 & WA & B & 8.75   & 6.19$\pm$0.17 &  -0.24$\pm$0.05 & 5.95$\pm$0.18 \\
314.073 & 0.304 & -1.994 & WA & C & 11.4   & 6.56$\pm$0.13 &  -0.39$\pm$0.04 & 6.17$\pm$0.14 \\
314.191 & 0.507 & -2.186 & WA & C & 8.57   & 6.37$\pm$0.12 &  -0.30$\pm$0.05 & 6.07$\pm$0.13 \\
315.100 & 0.411 & -1.890 & WA & A & 10.9   & 6.49$\pm$0.15 &  -0.36$\pm$0.04 & 6.13$\pm$0.15 \\
315.294 & 0.465 & -2.961 & RK & B & 5.86   & 6.50$\pm$0.12 &  -0.17$\pm$0.03 & 6.33$\pm$0.13 \\
316.482 & 1.107 & -1.267 & RK & C & 8.48   & 6.15$\pm$0.15 &  -0.14$\pm$0.05 & 6.01$\pm$0.16 \\
316.716 & 1.109 & -1.694 & GI & B & 7.37   & 6.36$\pm$0.14 &  -0.11$\pm$0.05 & 6.25$\pm$0.15 \\
316.986 & 0.838 & -1.851 & WA & B & 9.35   & 6.45$\pm$0.14 &  -0.22$\pm$0.05 & 6.23$\pm$0.15 \\
317.299 & 1.202 & -1.526 & WA & B & 7.60   & 6.19$\pm$0.19 &  -0.10$\pm$0.05 & 6.09$\pm$0.20 \\
317.319 & 0.842 & -1.885 & WA & C & 8.50   & 6.45$\pm$0.11 &  -0.20$\pm$0.05 & 6.25$\pm$0.13 \\
317.496 & 0.898 & -1.673 & RK & C & 9.25   & 6.25$\pm$0.14 &  -0.21$\pm$0.05 & 6.04$\pm$0.15 \\
317.530 & 1.204 & -1.545 & WA & C & 8.37   & 6.46$\pm$0.15 &  -0.12$\pm$0.05 & 6.34$\pm$0.17 \\
317.767 & 0.612 & -1.720 & RK & C & 11.0   & 6.49$\pm$0.16 &  -0.29$\pm$0.04 & 6.20$\pm$0.17 \\
318.047 & 0.961 & -1.822 & WA & B & 8.16   & 6.36$\pm$0.14 &  -0.16$\pm$0.05 & 6.20$\pm$0.15 \\
318.806 & 0.683 & -2.182 & WA & C & 6.56   & 6.15$\pm$0.14 &  -0.15$\pm$0.04 & 6.00$\pm$0.15 \\
318.931 & 1.032 & -1.840 & WA & A & 6.04   & 6.24$\pm$0.11 &  -0.08$\pm$0.03 & 6.15$\pm$0.12 \\
319.484 & 0.762 & -1.848 & WA & C & 8.76   & 6.36$\pm$0.15 &  -0.22$\pm$0.05 & 6.14$\pm$0.16 \\
319.910 & 0.735 & -2.209 & WA & C & 7.11   & 6.36$\pm$0.10 &  -0.17$\pm$0.04 & 6.19$\pm$0.12 \\
320.095 & 0.840 & -1.810 & WA & C & 9.64   & 6.46$\pm$0.23 &  -0.22$\pm$0.05 & 6.24$\pm$0.24 \\
320.623 & 0.786 & -2.180 & WA & B & 7.52   & 6.46$\pm$0.12 &  -0.17$\pm$0.05 & 6.29$\pm$0.13 \\
320.651 & 0.953 & -2.387 & WA & C & 5.19   & 6.51$\pm$0.11 &  -0.07$\pm$0.02 & 6.44$\pm$0.12 \\
320.677 & 1.252 & -1.804 & WA & B & 5.79   & 6.36$\pm$0.13 &  -0.05$\pm$0.03 & 6.31$\pm$0.14 \\
321.806 & 0.897 & -2.093 & WA & B & 7.27   & 6.46$\pm$0.11 &  -0.14$\pm$0.04 & 6.32$\pm$0.12 \\
322.336 & 1.016 & -1.828 & WA & B & 7.87   & 6.46$\pm$0.16 &  -0.14$\pm$0.05 & 6.32$\pm$0.17 \\
322.644 & 0.960 & -1.927 & RK & C & 6.67   & 6.23$\pm$0.15 &  -0.11$\pm$0.04 & 6.12$\pm$0.15 \\
323.514 & 1.836 & -1.164 & RK & C & 4.22   & 6.14$\pm$0.10 &   0.02$\pm$0.01 & 6.16$\pm$0.10 \\
323.757 & 1.016 & -2.456 & RK & B & 3.74   & 6.36$\pm$0.11 &  -0.06$\pm$0.01 & 6.30$\pm$0.11 \\
323.855 & 1.031 & -2.104 & WA & C & 5.14   & 6.34$\pm$0.13 &  -0.06$\pm$0.02 & 6.28$\pm$0.13 \\                                      
324.144 & 1.199 & -1.803 & WA & B & 6.67   & 6.36$\pm$0.14 &  -0.07$\pm$0.04 & 6.29$\pm$0.15 \\
324.281 & 1.022 & -2.540 & RK & C & 3.79   & 6.45$\pm$0.11 &  -0.06$\pm$0.01 & 6.39$\pm$0.11 \\
324.915 & 1.059 & -2.482 & RK & C & 5.08   & 6.59$\pm$0.13 &  -0.06$\pm$0.02 & 6.53$\pm$0.13 \\
325.259 & 1.297 & -1.807 & WA & C & 5.55   & 6.36$\pm$0.12 &  -0.05$\pm$0.02 & 6.32$\pm$0.13 \\
325.549 & 1.300 & -1.829 & WA & A & 5.34   & 6.35$\pm$0.10 &  -0.03$\pm$0.02 & 6.32$\pm$0.11 \\
326.106 & 1.788 & -1.427 & RK & C & 2.23   & 5.96$\pm$0.14 &   0.00$\pm$0.00 & 5.96$\pm$0.14 \\
326.311 & 1.146 & -2.387 & RK & C & 3.85   & 6.43$\pm$0.11 &  -0.05$\pm$0.01 & 6.38$\pm$0.11 \\
326.665 & 1.191 & -2.296 & RK & C & 3.45   & 6.35$\pm$0.11 &  -0.04$\pm$0.00 & 6.30$\pm$0.11 \\
326.904 & 1.566 & -1.846 & RK & C & 1.95   & 6.03$\pm$0.14 &  -0.02$\pm$0.00 & 6.00$\pm$0.14 \\
327.073 & 1.195 & -2.346 & RK & B & 2.74   & 6.26$\pm$0.15 &  -0.05$\pm$0.00 & 6.21$\pm$0.15 \\
327.420 & 1.331 & -1.810 & RK & C & 4.70   & 6.24$\pm$0.11 &  -0.03$\pm$0.01 & 6.21$\pm$0.11 \\
329.126 & 1.488 & -2.479 & RK & C & 1.31   & 6.36$\pm$0.15 &  -0.04$\pm$0.00 & 6.32$\pm$0.15 \\   
329.168 & 1.612 & -1.490 & RK & C & 3.37   & 6.04$\pm$0.11 &  -0.01$\pm$0.00 & 6.03$\pm$0.11 \\
\hline                                                       
\end{supertabular}}
\twocolumn

\section{Determination of the S/N ratio in the observed spectrum of HD~122563\label{SN-ratio}}

\begin{figure}[!t]
	\centering
	\includegraphics[width=\columnwidth]{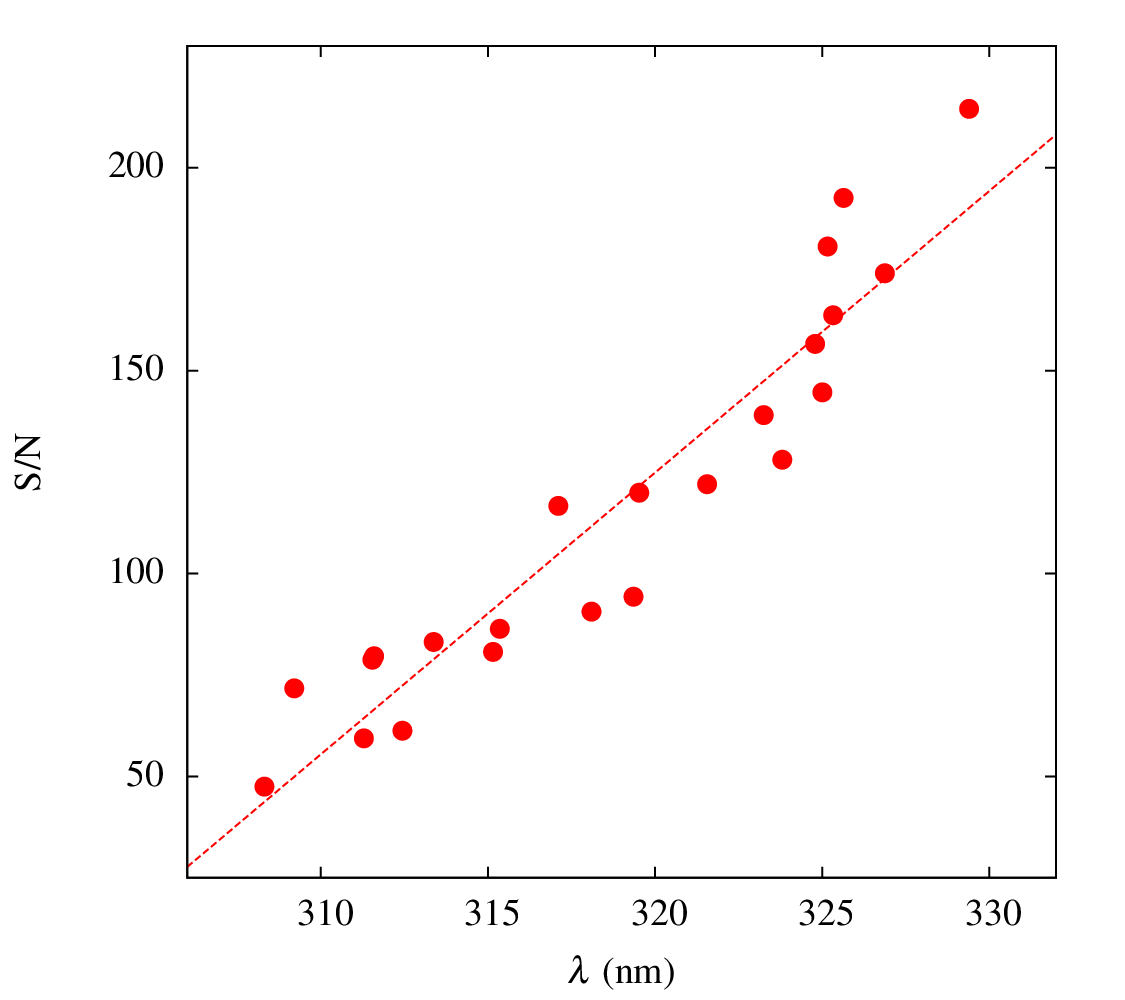}
	\caption{Signal-to-noise ratio in the investigated spectral region. The dashed line indicates a linear fit to the data.}
	\label{fig:sn}
\end{figure}

In Eq.~(\ref{eq:fitbyline}), $\sigma_{\rm flux}$ represents the photon noise, which is the inverse of the $S/N$ ratio. $\sigma_{\rm flux}$ was estimated as a standard deviation of flux values in a blend-free (continuum) spectral regions. Measurements were done in the whole investigated spectral range and 10-20 wavelength points were used for each measurement of $\sigma_{\rm flux}$. The results are shown in Fig.~\ref{fig:sn}. A rather tight relationship between $S/N$ and $\lambda$ was found and for further applications it is approximated with a linear fit.

\section{Microturbulence in 3D model atmospheres\label{app:vmic}}
           
\begin{figure}[!h]
	\centering
	\includegraphics[width=\columnwidth]{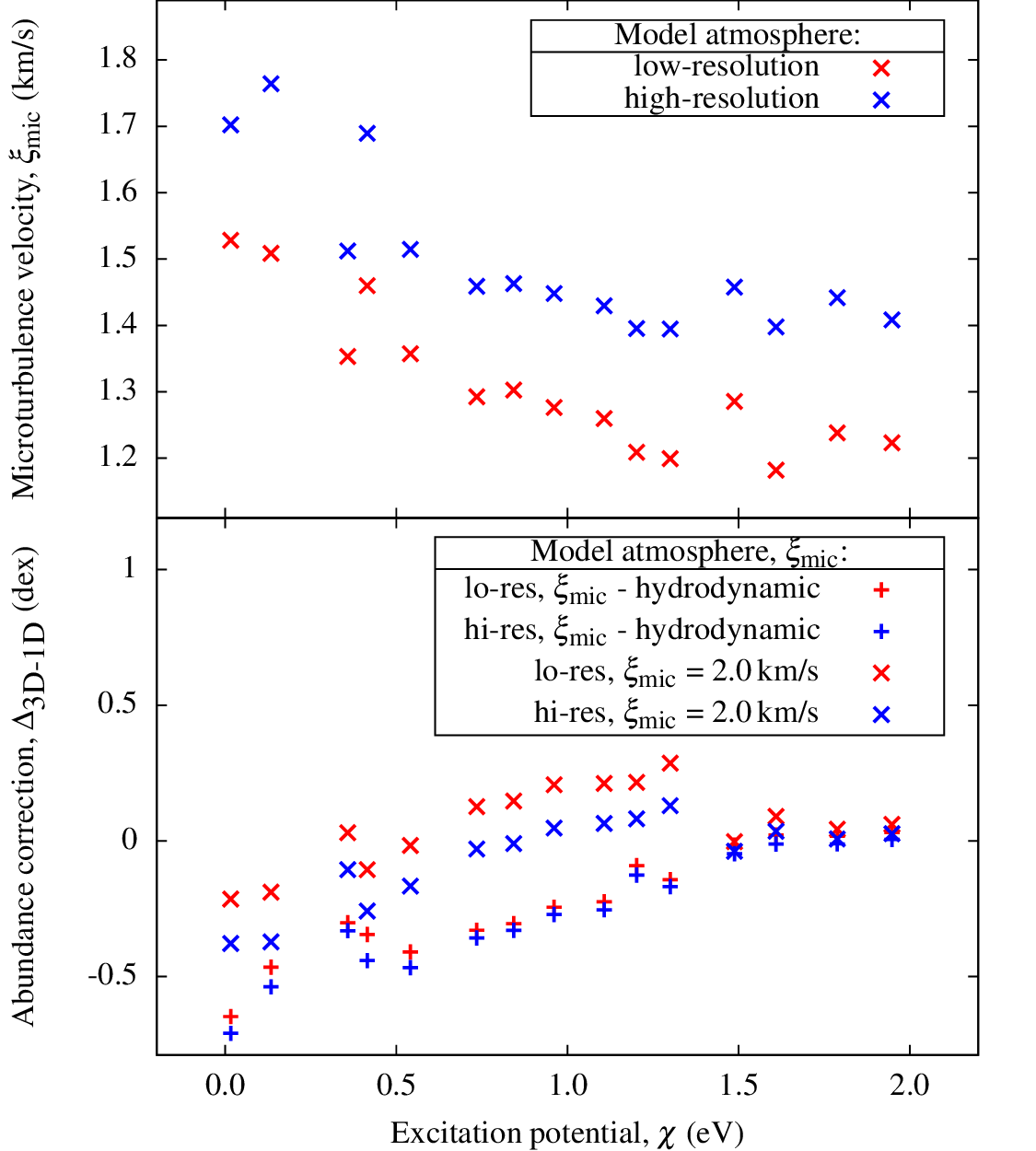}
	\caption{\textbf{Top:} microturbulence velocities, \vmic, determined from the model atmospheres that were computed using different spatial resolution. \textbf{Bottom:} 3D--1D abundance corrections computed with the two model atmospheres and different approach to \vmic\ (see text and legend for details) . }
	\label{fig:micres}
\end{figure}	

    We have argued in Sect.~\ref{sect:3D_abund_method} that, in order to perform a strictly differential 3D--1D analysis of strong lines that are sensitive to the microturbulence, 3D--1D abundance corrections have to be computed using \vmic\ derived from the corresponding 3D hydrodynamical model atmosphere. In this Appendix we justify our claims.

\begin{table}[!h]
	\centering
	\caption{Spatial characteristics of the model atmospheres of HD~122563 used and respective mean \vmic\ values derived in Appendix~\ref{app:vmic}.}\label{tab:modelsvmic}
\begin{tabular}{|c|c|c|c|}
\hline 
Spatial     & Spatial extent & Grid points         &  $\xtmean{\mbox{\vmic}}$  \\ 
resolution  &   [Mm]         & $x\times y\times z$ &  km/s     \\
\hline 
Low  & $4.38^{2}\times2.90$ & $160^{2}\times200$   & $1.31 \pm 0.11$ \\ 
\hline 
High & $4.40^{2}\times3.13$ & $375^{2}\times330$   & $1.50 \pm 0.12$ \\ 
\hline 
\end{tabular} 
\end{table}

    We have used two 3D model atmospheres of different spatial resolution and very similar spatial extent in order to compute \vmic\ values using the procedure described in Sect.~\ref{sect:3D_abund_method} (see Table~\ref{tab:modelsvmic} for a description of the model atmospheres and results). The results are shown in Fig.~\ref{fig:micres} (Top). One may note immediately that the high-resolution model atmosphere predicts higher \vmic\ values. This is expected since higher spatial resolution allows for the appearance of smaller-scale turbulent eddies \citep[e.g.,][]{SCL13}. Then, for each of the two models we computed a mean microturbulence velocity by averaging \vmic\ values determined using 15 OH UV lines. These two values (cf. Table~\ref{tab:modelsvmic}) of the mean microturbulence velocity were used to compute 3D--1D abundance corrections corresponding to the low- and high-resolution cases (Fig.~\ref{fig:micres}, Bottom). The obtained results show that, once \vmic\ is calibrated with the help of the 3D model atmosphere, the resulting abundance corrections are very similar. This is because in both the low- and high-resolution case \vmic\ was adjusted in such a way that the 1D model atmospheres would reproduce the line strengths predicted with the corresponding 3D hydrodynamical model atmosphere as accurately as possible. This, in effect, made the 3D--1D abundance corrections insensitive to both the shortcomings of the 3D hydrodynamical model atmospheres (i.e., spatial resolution in our case) and microturbulence used in the 1D~LTE abundance analysis. On the other hand, if the 3D--1D abundance corrections were computed using some fixed value of \vmic\ (for example, the one determined in the process of 1D~LTE abundance analysis), the abundance corrections obtained using the low- and high-resolution 3D hydrodynamical model atmospheres will be different. This is illustrated in Fig.~\ref{fig:micres} (Bottom) where we also show 3D--1D abundance corrections computed using the two 3D hydrodynamical model atmospheres and identical $\vmic = 2.0$\,km/s utilized with the 1D \ATLAS\ model atmospheres. Indeed, in this case the differences in the abundance corrections are significant and reach $\sim 0.2$\,dex.
  
\section{Importance of the C/O ratio in the 3D spectral synthesis\label{app:COratio}}

\begin{figure}[!t]
 \centering
 \includegraphics[width=\columnwidth]{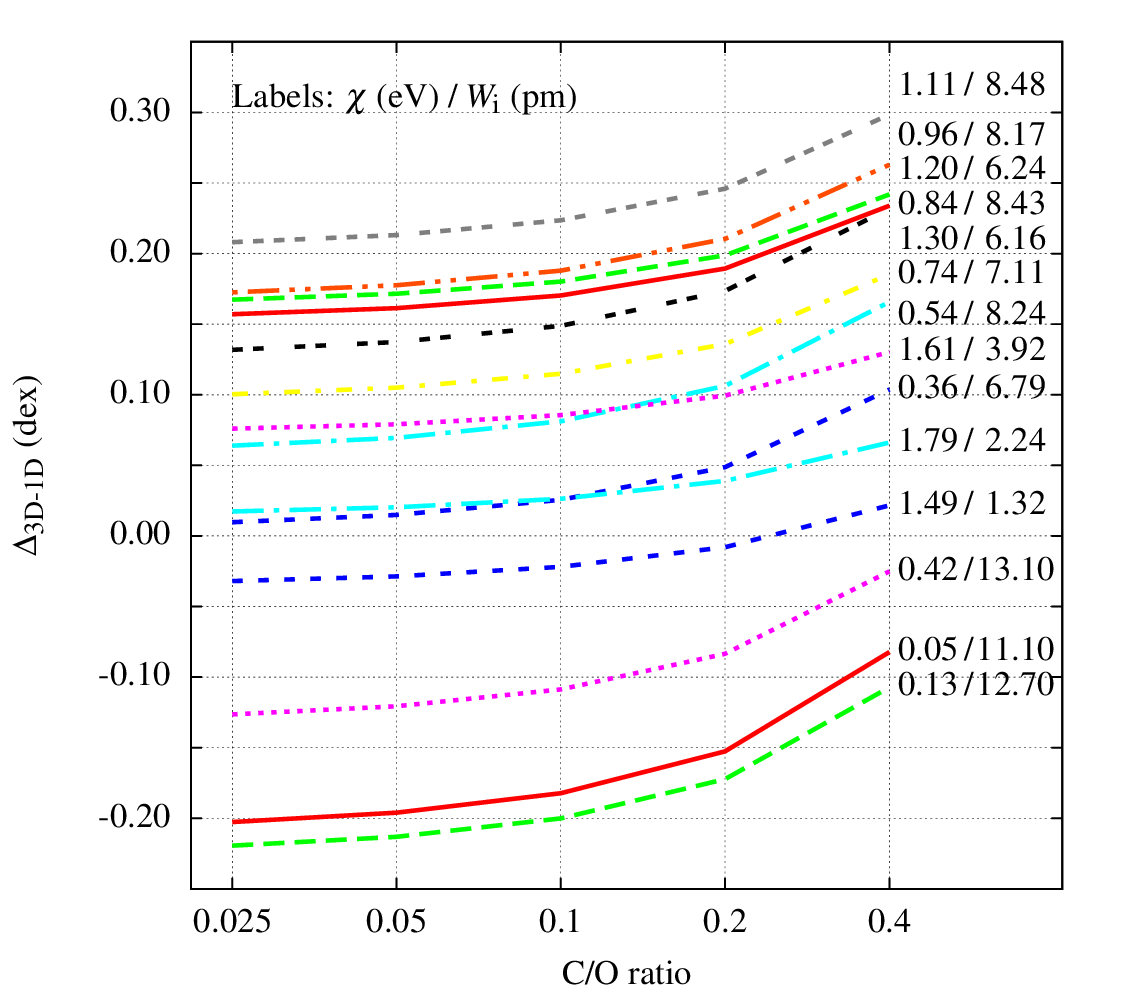}
 \caption{Dependency of $\Delta_{\rm 3D-1D}$ of a sample of 14 OH UV lines on C/O ratio. Labels on the right indicate $\chi /$eV and $\Wi/$pm of individual lines.}
 \label{fig:CO}
\end{figure}

    As recently noted by \citet{GCB16}, formation of carbon- and oxygen- bearing molecules are very sensitive to the C/O abundance ratio and abundances of these elements should be determined simultaneously \citep[see also][]{JGH10}. The C/O ratio determines which chemical species -- carbon or oxygen -- is in minority and which of these chemical species will be locked by the formation of strong CO molecules \citep[see][who, in their Appendix~C, investigated case where ${\rm C/O} < 1$]{DKS13}. Moreover, line formation in 3D model atmospheres is expected to be more sensitive to the C/O ratio than that of 1D model atmospheres, hence we cannot expect that the possible errors would cancel in our differential approach.
   
    In case of HD~122563, a 1D-based analysis of \citep{SCP05} suggests $A{\rm (C)} = 5.23$ and $A{\rm (O)} = 6.54$, what results in a very low ${\rm C/O} \approx 0.05$. \citep{AM15} used LTE analysis of molecular lines and NLTE analysis of atomic lines and derived $A{\rm (C)} = 5.10$, which implies even lower ${\rm C/O} \approx 0.03$. Also, such C/O ratio means that oxygen is strongly majority species, variation of $A{\rm (C)}$ should not influence formation of OH lines and we should be safe with using 1D-based $A{\rm (C)}$. However, we have carried out tests that strengthen our claims.
      
    Fig.~\ref{fig:CO} shows 3D-1D abundance corrections plotted versus C/O ratio for a set of 14 real (i.e. real $\chi$ and measured \Wi ) OH lines. It is seen that when $0.025 < \mathrm{C/O} < 0.1$ (nearly, 0.6\,dex in $A{\rm (C)}$) the abundance corrections of most lines vary by less than 0.01\,dex. We hence conclude that uncertainties on the C/O ratio does not influence our results in any significant way and 1D-based $A{\rm (C)}$ could be used without introducing additional uncertainties. We add that this could be expected for the majority of red giant branch stars because such stars have experienced the first dredge-up and had their carbon depleted relative to the initial value.

   \section{The influence of 3D hydrodynamical effects on the formation of OH UV lines and 3D--1D oxygen abundance corrections\label{app:abucorr_add}}
   
    \begin{figure}[!t]
   	\centering               
   	\includegraphics[width=\columnwidth]{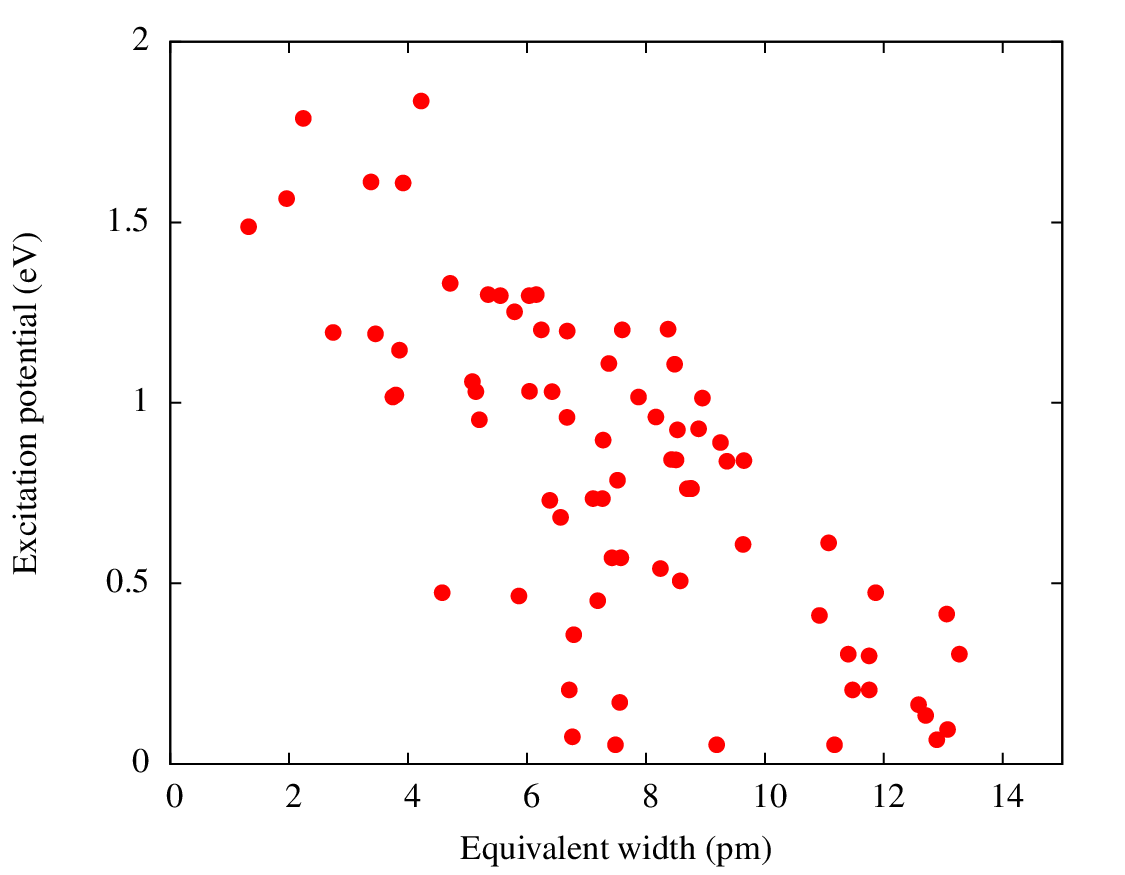}
   	\caption{Line excitation potential, $\chi$, of OH UV lines used in our study plotted versus the line equivalent width, \Wi.}
   	\label{fig:Wchi}
   \end{figure}

   \begin{figure}[!t]
   \centering
   \includegraphics[width=\columnwidth]{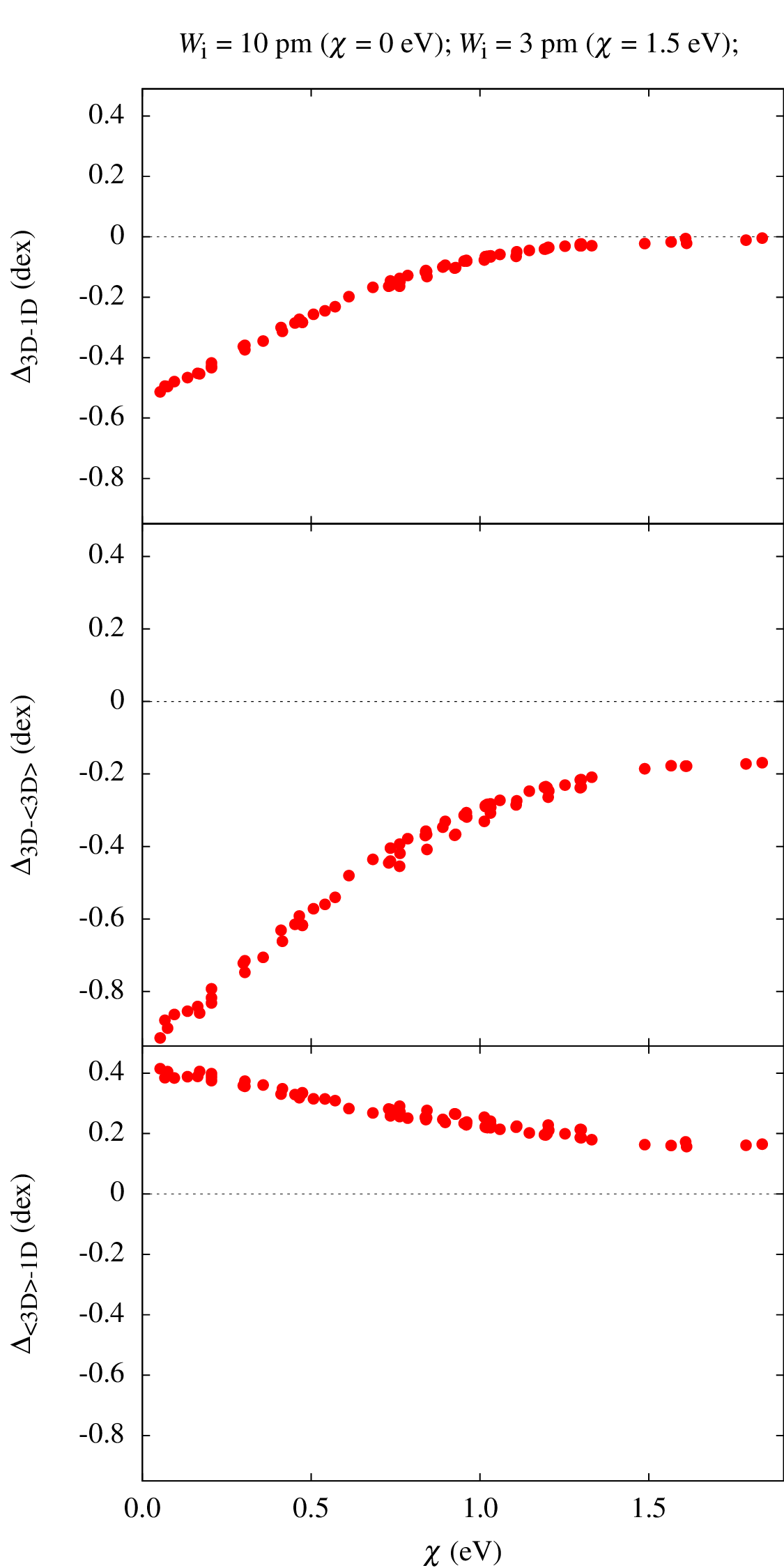}
   \caption{The three abundance corrections plotted against excitation potential $\chi$. Abundance corrections were computed at line strength linearly varying with the excitation potential where $\Wi = 10$\,pm for lines with $\chi = 0$\,eV and $\Wi = 3$\,pm for lines with $\chi = 1.5$\,eV.}
   \label{fig:dabu_add_varW}
   \end{figure}
   
   Before going into the investigation of 3D--1D oxygen abundance corrections, let us first note that there is a relation between the line excitation potential, $\chi$, and line equivalent width, \Wi\ (Fig.~\ref{fig:Wchi}). As we will show below, this has a direct influence on the 3D--1D abundance corrections for OH UV lines discussed in Sect.~\ref{sect:3D_abund_method}.

   \begin{figure*}[!ht]
   \centering
   \includegraphics[width=13cm]{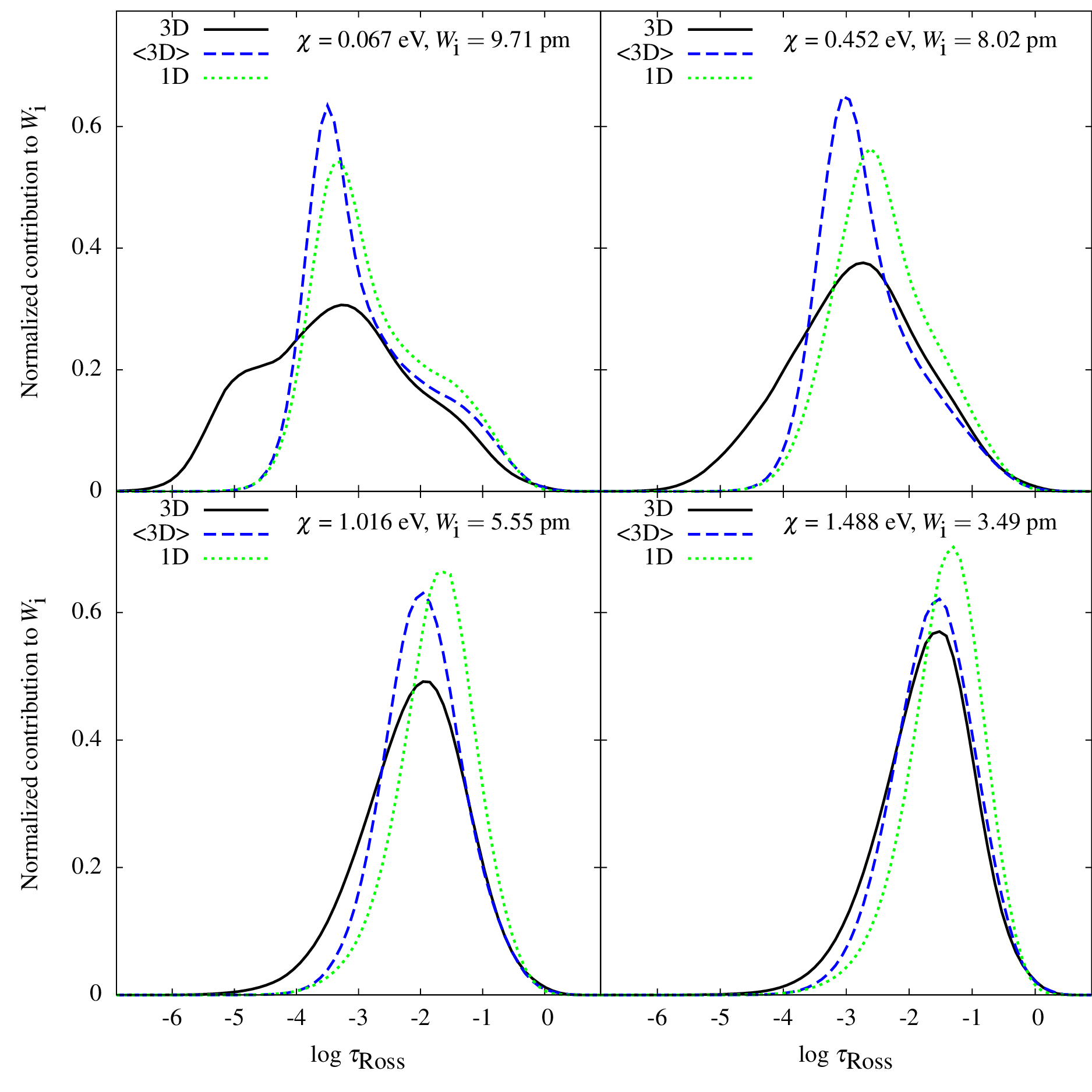}
   \caption{3D and 1D contribution function to equivalent width for the four lines that are defined by a linear $\chi$ - \Wi\ relationship. See Table~\ref{tab:abucorr_varW} and text for details.}
   \label{fig:contf_warW}
   \end{figure*}

   \begin{table}[!h]
   \caption{Line parameters and abundance corrections -- $\Delta_{\rm 3D-1D}$, $\Delta_{\rm 3D-\langle3D\rangle}$, $\Delta_{\rm \langle3D\rangle-1D}$ -- for the four lines displayed in Fig.~\ref{fig:contf_warW}. These lines are defined by a fictitious \Wi\ that is determined by an approximate linear fit to a $\chi$ - \Wi\ relationship in Fig.~\ref{fig:Wchi}}\label{tab:abucorr_varW}
   \centering
   
   \begin{tabular}{|c|c|c|c|c|c|}
   \hline 
   $\lambda$ & $\chi$ & \Wi & $\Delta_{\rm 3D-1D}$ & $\Delta_{\rm 3D-\langle3D\rangle}$ & $\Delta_{\rm \langle3D\rangle-1D}$\\ 
   nm        & eV     & pm  & dex & dex & dex \\ 
   \hline 
   310.123 & 0.067    & 9.71 & -0.50 & -0.88 & 0.38 \\
   313.618 & 0.452    & 8.02 & -0.29 & -0.62 & 0.33 \\
   322.337 & 1.016    & 5.55 & -0.07 & -0.29 & 0.22 \\
   329.126 & 1.488    & 3.49 & -0.02 & -0.19 & 0.17 \\
   \hline 
   \end{tabular} 
   \end{table}
      
   To understand the basic tendencies in the behaviour of 3D--1D abundance corrections seen in Fig.~\ref{fig:dabu_main}, we computed abundance corrections for spectral lines with variable line excitation potential, $\chi$, and equivalent width, \Wi. The relation between $\chi$ and $\Wi$ was defined in such a way that it would approximately follow the best-fitting line in Fig.~\ref{fig:Wchi}. The resulting abundance corrections are shown in Fig.~\ref{fig:dabu_add_varW} while the contribution functions for four selected lines are shown in Fig.~\ref{fig:contf_warW}. These figures together give a clear indication that the relation between $\chi$ and $\Wi$ in our OH UV line sample plays an important role in defining the 3D--1D abundance corrections. At the lowest excitation potentials lines are strongest and thus their formation reaches the outermost atmospheric layers where temperature and line opacity fluctuations are largest. This leads to large negative $\Delta_{\rm 3D-\langle3D\rangle}$ corrections which are counterweighted to some extent by significant but positive $\Delta_{\rm \langle3D\rangle - 1D}$ abundance corrections. With increasing line excitation potential, lines also become weaker and thus their formation is confined to deeper atmospheric layers where temperature fluctuations are smaller. This leads to smaller (i.e., more positive) $\Delta_{\rm 3D-\langle3D\rangle}$ corrections\footnote{Note that when the line excitation potential is in the range of $0 < \chi \la 1$\,eV, the line equivalent widths are equal to $\approx10-5$\,pm and thus the line strength is sensitive to the velocity field. As shown in Fig.~\ref{fig:vmic}, velocity field in the 3D model atmosphere also diminishes with $\chi$ which further weakens spectral lines in 3D.}. Since the depth of the line formation changes insignificantly with further increasing line excitation potential, the $\Delta_{\rm 3D-\langle3D\rangle}$ correction stops growing with $\chi$ and the behaviour of the total abundance correction, $\Delta_{\rm 3D-1D}$ is governed by the slowly decreasing $\Delta_{\rm \langle3D\rangle - 1D}$ correction.

\end{appendix}

\end{document}